\documentclass{aa}
\usepackage{graphicx}
\usepackage{verbatim}
\usepackage{textcomp}
\usepackage{color}
\usepackage{afterpage}
\usepackage{longtable}
\usepackage{enumerate}
\usepackage{multirow}
\usepackage{pdflscape}
\usepackage[square,authoryear]{natbib}
\usepackage[margin=0.7in]{geometry} 
\bibpunct{(}{)}{;}{a}{}{,} 
\usepackage{verbatim}
\usepackage{microtype}
\usepackage{fixltx2e} 
\usepackage{amsmath}
\usepackage{amssymb}

\title{Low frequency measurements of synchrotron absorbing \\ HII regions and modeling of observed synchrotron emissivity }
\titlerunning{Synchrotron emissivities, catalog and modeling}
\author{I.M. Polderman \inst{\ref{inst1}}
\and M. Haverkorn \inst{\ref{inst1}}
\and T.R. Jaffe \inst{\ref{inst2}, \ref{inst3}}
\and M.I.R. Alves \inst{\ref{inst1}} }

\authorrunning{Polderman I.M. et al.}
\institute{Department of Astrophysics/IMAPP, Radboud University, P.O. Box 9010, 6500 GL Nijmegen, The Netherlands.\label{inst1}
\and 
CRESST II, NASA Goddard Space Flight Center, Greenbelt, MD, 20771, USA.\label{inst2}
\and
Department of Astronomy, University of Maryland, College Park, MD, 20742, USA.\label{inst3}
}

\abstract
{Cosmic rays (CRs) and magnetic fields are dynamically important components in the Galaxy, and their energy densities are comparable to that of the turbulent interstellar gas. The interaction of CRs and Galactic magnetic fields produces synchrotron radiation clearly visible in the radio regime. Detailed measurements of synchrotron radiation averaged over the line-of-sight (LOS), so-called synchrotron emissivities, can be used as a tracer of the CR density and Galactic magnetic field (GMF) strength.  }
{Our aim is to model the synchrotron emissivity in the Milky Way using a 3 dimensional dataset instead of LOS-integrated intensity maps on the sky.}
{Using absorbed HII regions we can measure the synchrotron emissivity over a part of the LOS through the Galaxy, changing from a 2 dimensional to a 3 dimensional view. Performing these measurements on a large scale is one of the new applications of the window opened by current low frequency arrays. Using various simple axisymmetric emissivity models and a number of GMF-based emissivity models we can simulate the synchrotron emissivities and compare them to the observed values in the catalog.}
{We present a catalog of low-frequency absorption measurements of HII regions, their distances and electron temperatures, compiled from literature. These data show that the axisymmetric emissivity models are not complex enough, but the GMF-based emissivity models deliver a reasonable fit. These models suggest that the fit can be improved by either an enhanced synchrotron emissivity in the outer reaches of the Milky Way, or an emissivity drop near the Galactic center.}
{State-of-the-art GMF models plus a constant CR density model cannot explain low-frequency absorption measurements, but the fits improved with slight (ad-hoc) adaptations. It is clear that more detailed models are needed, but the current results are very promising.}
\keywords{ISM: cosmic rays -- ISM: Galactic magnetic fields -- ISM: HII regions -- Galaxy: structure -- Radio continuum: ISM -- catalogs}

\begin{document}
\maketitle
\section{Introduction}
\label{sec:intro}

Cosmic rays (CRs) play an important role in many aspects of Galaxy ecology \citep{Grenier2015}. Consequently, understanding more about these particles is a big leap forward in understanding our own Galaxy. 
In general CRs moderate processes by depositing or taking away energy. For example the CRs that are accelerated in a supernova remnant (SNR) take away between 10\% and 50\% of the SNR energy. This lowers the temperature of the remnant and changes the impact of the SNR on the interstellar medium (ISM). The presence of relativistic particles also causes a longer expansion time for the SNR, thereby increasing the momentum that can be deposited in the ISM \citep{Diesing2018}. 
While CRs diffuse through the Galaxy, their interactions with the ISM cause them to deposit energy along kiloparsecs. The interactions can cause ionization and heating of dense interstellar gas, thereby keeping the darkest ISM at temperatures near 10~K, for example. This also makes CRs very important for driving chemistry at low temperatures. 
CRs are also thought to drive magnetohydrodynamic waves, which could maintain interstellar turbulence, and large-scale interstellar flows. Among these flows are Galactic winds, fountains and possibly bubbles -- like the Fermi bubbles, \citep[e.g.,][]{Mertsch2017,SuM2010}. Furthermore, CRs probe objects and environments in the Galaxy, like stellar atmospheres, the composition of accelerated matter, and the total gas content of interstellar clouds. CRs also interact with the Galactic magnetic field (GMF). 
This interaction produces the synchrotron radiation that dominates the radio spectrum at low radio frequencies. At these frequencies, the synchrotron intensity can be used as a measure for both the CR number density, $n_{\rm CR}$, and the strength of the magnetic field, $B_{\bot}$, perpendicular to the the line-of-sight (LOS) \citep{Webber1977,Jaffe2010} 
\begin{eqnarray}
I_{synch} &=& \underset{L}{\int} n_{CR}~B_{\bot}^{\frac{p+1}{2}} ~dL
\end{eqnarray} 
The value of $p$ in the exponent of the magnetic field is assumed to be 3, which follows from the CR spectrum, \citep{Planck2016}.

One complication of this tracer however, is the averaging over the entire LOS. It is unlikely that the synchrotron emission is constant along the LOS. To partially overcome this issue the low-frequency regime opens up the possibility to integrate over only part of it. To accomplish this, observations of HII regions with known distances are used, essentially elevating this kind of work from a 2-dimensional view (total LOS) of Galactic synchrotron emission to a 3-dimensional one (partial LOS).

\citet{Scheuer1953} were among the first to consider the HII region effect on synchrotron, or non-thermal, radiation. And \citet{Shain1959} showed that at low frequencies free-free absorption starts to play an important role. This absorption is characteristic for an optically thick HII region, where the free-free opacity, $\tau$ $>$ 1, which is only the case at lower frequencies ($\nu \lesssim$ 150\,MHz~\citep{Hindson2016}). Several papers are dedicated to determining the spectral turn-over frequency for the free-free emission of Galactic HII regions \citep[e.g.,][]{Hindson2016,Mezger1967}, below which the opacity increases and HII regions are observed as absorption regions against a Galactic synchrotron background emission. As will be explained in detail in Sect.~\ref{sec:theory}, using low-frequency observations of HII regions allows one to integrate over only a section of the LOS. Several papers \citep[e.g.,][]{Jones1974,Caswell1976,Roger1999,Nord2006,Hindson2016,Su2016} determine synchrotron emissivity values using this method. These observations can be used to gain insight into CR density and magnetic field strength in the Milky Way, and different models can be used to describe the observed values \citep{Nord2006,Su2016}. In this paper these models are known as the simple axisymmetric emissivity models.

Another complication of synchrotron as a tracer is the fact that it is a convolved effect of both the CR density and the magnetic field strength. It is clear that CR particles are strongly coupled to magnetic fields, and a common used simplification is the assumption of equipartition \citep[e.g.,][]{Beck2015,Ferriere2016} between the energy densities of the CRs and the magnetic field. This assumption is plausible on large scales, but less likely on scales of (tens of) parsecs. If we want to draw any conclusions about CR density and GMF strength in the Milky Way we will need an assumption like equipartition. In this work we also try another approach, by using existing GMF models and an assumed CR density model to compute synchrotron emissivity. In this paper the combined GMF and CR density models, are known as the GMF-based emissivity models. The variations in the emission are based solely on the (different) details in the GMF models. By comparing the simulated and observed emission we can start understanding the differences between them and between the GMF models.

This paper is constructed as follows: Sect.~\ref{sec:theory} discusses Galactic absorption features and extracting synchrotron emissivity. Section~\ref{sec:data} discusses the data we have collected from HII region observations, and Sect.~\ref{sec:catalog} describes the effort of combining these data into one coherent master catalog. Section~\ref{sec:analysis} discusses the features in the observed emissivities. Section~\ref{sec:modeling} describes the modeling effort and its results are discussed in Sect.~\ref{sec:results}. Section~\ref{sec:discussion} contains the discussion of this work and Sect.~\ref{sec:conclusion} the conclusions. Section~\ref{sec:futurework} discusses future work.

\section{Galactic absorption features and synchrotron emissivity}
\label{sec:theory}

We explain the interpretation of measurements of Galactic free-free absorption.
For low radio frequency observations of HII regions, with an absolute calibration of the flux, the observed brightness temperature T$_{\rm obs}$ is \citep{Kassim1990}
\begin{eqnarray}
T_{\rm obs} = T_{\rm F} +T_{\rm e} \,(1 - e^{-\tau})+T_{\rm B}\,e^{-\tau},
\label{eq:kassimfull}
\end{eqnarray}
where $T_{\rm F}$ is the brightness temperature of the synchrotron emission between the observer and the HII region (foreground) and $T_{\rm B}$ is the brightness temperature of the synchrotron emission behind the HII region (background). The HII region is described by the electron temperature, $T_{\rm e}$, and $\tau$, the opacity. When observing at low frequencies, $\tau\,\gg\,$ 1 so the equation reduces to
\begin{eqnarray}
T_{\rm obs} = T_{\rm e} + T_{\rm F}.
\label{eq:approxsd}
\end{eqnarray}

For observations without absolute calibration (usually the case for interferometric data), the large scale flux is unobserved, which can be envisioned as subtracting the total temperature $T_{\rm T} = T_{\rm F} + T_{\rm B}$ to obtain
\begin{eqnarray}
T_{\rm obs} &=& T_{\rm F} +T_{\rm e}\, (1 - e^{-\tau})+T_{\rm B}\, e^{-\tau} - T_{\rm T} \\
 &=& (T_{\rm e} - T_{\rm B})\,(1 - e^{-\tau}).
\label{eq:kassiminterf}
\end{eqnarray} 

Using the high-opacity approximation for the low frequencies, we get
\begin{eqnarray}
T_{\rm obs} = T_{\rm e} - T_{\rm B}.
\label{eq:approxint}
\end{eqnarray}

At low frequencies $T_{\rm B} > T_{\rm e}$ is generally valid, so that $T_{\rm obs} < 0$. The electron temperature can be estimated or it can be derived from higher frequency observations \citep[e.g.][]{Paladini2004,Alves2012,Balser2015}. 

Depending on the type of measurement (single dish or interferometer) and the calibration (absolute or not), we obtain the emissivity $\epsilon$ in either the foreground or the background of an HII region as 

\begin{eqnarray}
\epsilon_{\rm F}&=&T_{\rm F}/D_{\rm F}\\
\epsilon_{\rm B}&=&T_{\rm B}/D_{\rm B},
\label{eq:emiss}
\end{eqnarray}

where $D_{\rm F}$ is the distance to the HII region and $D_{\rm B}$ is the distance from the HII region to the edge of the synchrotron disk, along the LOS. For both of these distances we will use the term path length. We assume a Galactic disk with a radius of 20 kpc and a solar distance to the Galactic center of 8.5 kpc. Usually HII region distances are determined kinematically, for nearby HII regions it can be done by determining the spectral type of the ionizing star. Another method uses the association with a maser with a measured parallax \citep{Anderson2014}.

The correct foreground and background temperatures can only be achieved if the opacity $\tau \gg$ 1 across the observing beam. Therefore we consider only HII regions larger than the beam.


\section{Data}
\label{sec:data}

We use data from five published papers to build our catalog. Relevant details of these papers are presented in Table~\ref{table:sourcepapersinfo}. We discuss these papers by the number assigned to them in this table.
\subsection{Background on papers used}
\label{sec:datapapers}

\begin{table*}
\caption{Source paper observing parameters}              
\label{table:sourcepapersinfo}      
\centering                                      
\begin{tabular}{*{4}{c}rrcllc}          
\hline\hline                        
\# & Authors & Telescope & $\nu$ (MHz) & \multicolumn{5}{c}{Range} & Resolution \\    
\hline                                   
    1 & \citet{Jones1974} & Fleurs N.S.W. & 29.9  & 225$^{\circ}$& \textless& $\ell$ &\textless& 30$^{\circ}$ &0.8$^{\circ}$ at zenith  \\      
    2 & \citet{Roger1999} & DRAO      & 22  &$-$28$^{\circ}$&\textless& decl.& \textless& +80$^{\circ}$ & 1.1$^{\circ}$ $\times$ 1.7$^{\circ}$ \\
    3 & \citet{Nord2006} & VLA        &74  & 26$^{\circ}$& \textless& $\ell$& \textless& $-$15$^{\circ}$ &  7.33' $\times$4.75'\\
&&&&&&&&&6.67'$\times$5.67'\\
&&&&&&&&&17.41'$\times$6.67'\\
&&&&&&&&&8.33'$\times$5.00'\\
    4 & \citet{Hindson2016} & MWA      & 88 & +25$^{\circ}$ &\textless& decl. &\textless &$-$9$^{\circ}$ &5.6' $\times$ 5.1' \\
    5 & \citet{Su2016} & MWA      & 88  & 250$^{\circ}$ &\textless& $\ell$& \textless &355$^{\circ}$ & 5.6' $\times$ 5.6'\\
\hline                                             
\end{tabular}
\end{table*}

\citet{Jones1974}, hereafter paper 1, map brightness temperatures near the Galactic plane between Galactic longitudes $\ell$ = 225$^{\circ}$ and $\ell$ = 30$^{\circ}$ from a survey with the telescope in Fleurs, N.S.W. \citep{Finlay1973} at 29.9 MHz. These authors observe 29 absorption features of which 2 are optically thick across the beam. 
The absolute calibration for this data, by using a 30 MHz Parkes survey \citep{Mathewson1965} among others, results in foreground synchrotron emissivities for the HII regions from this paper. 

\citet{Roger1999}, hereafter paper 2, uses data taken with the 22 MHz telescope at the Dominion Radio Astrophysical Observatory (DRAO), in Penticton B.C. Canada \citep{Costain1969}, in the period between 1965 and 1969. They map the 22 MHz emission between $\ell$ = $-28^{\circ}$ and $\ell$ = +80$^{\circ}$ . These observations include 21 HII regions of which 8 were larger than the beam. We discard G6.3+26.5 because of its high latitude, its location in the direction of the North Polar Spur and its anomalously high observed emissivity.
These authors performed an absolute calibration of their data by considering the average brightness temperature ratio between their data and data from the absolute calibrated 408 MHz all--sky survey by \citet{Haslam1982}. Their contribution to this catalog consists of foreground emissivities only. 

\citet{Nord2006}, hereafter paper 3, observed 4 fields along the Galactic plane between $\ell$ = 349$^{\circ}$ and $\ell$ = 21.5$^{\circ}$ with the VLA in compact configuration at 74 MHz. Their goal was a dedicated search for absorption features in the Galactic plane. The observations include 92 HII regions of which 41 had known distances and thus, are usable. 
To identify HII regions, paper 3 used an a priori search for HII absorption features, at locations of known HII regions from \citet{Paladini2003}. 
This paper has done no absolute calibration, therefore the emissivities obtained are background emissivities.

\citet{Hindson2016}, hereafter paper 4, uses data taken as part of the Galactic and Extragalactic All--sky MWA survey, (GLEAM, \citealt{Hurley-Walker2017}). This survey uses multiple frequency bands between 72 and 231 MHz. The authors of paper 4 make use of this wide frequency range to identify the characteristic spectrum generated by the thermal Brehmsstrahlung process in HII regions. Large scale background emission observed by MWA is subtracted so that Eq.~\ref{eq:approxint} holds. Out of the 302 detected HII regions 29 have known distances. A further 2 are discarded because they are smaller than the beam.     

\citet{Su2016}, hereafter paper 5 use part of the data from the GLEAM survey at 88 MHz, between $\ell$ = 250$^{\circ}$ and $\ell$ = 355$^{\circ}$, to measure free--free absorption of the Galactic synchrotron emission by intervening HII regions along the LOS. They perform a dedicated search for absorption regions, at locations of known HII regions from \citet{Anderson2014}, and find 47 that are usable. Like paper 4 they subtract the large scale background emission.

\begin{table}
\caption{Source paper catalog contributions}              
\label{table:sourceentriesinfo}      
\centering                                      
\begin{tabular}{*{3}{c}}          
\hline\hline                        
\# & Back- or Foreground & Entries \\    
\hline                                   
    1 & Foreground& 2    \\      
    2 &Foreground & 7      \\
    3 & Background & 41         \\
    4 & Background & 27      \\
    5 & Background & 47      \\
\hline                                             
\end{tabular}
\end{table}
 
\subsection{Flux Calibration}
\label{sec:flux}

Since some of the data used are over 50 years old, we checked that all data are calibrated to the same standardized flux levels. 
Paper 1 uses Hydra A -- also known as 3C218 -- as a calibrator, with a flux at 29.9 MHz measured as 1512\,$\pm$\,81\,Jy \citep{Finlay1973}.
Both papers 2 and 3 use Cygnus A as a calibrator. Paper 2 uses a flux for Cygnus A of 29100\,$\pm$\,1500\,Jy at 22\,MHz \citep{Roger1969}. Paper 3 uses the \citet{Baars1977} flux scale of 1706\,$\pm$\,50\,Jy at 74\,MHz, which also states a value of 29100\,$\pm$\,1750\,Jy at 22\,MHz. These two values for the flux of Cygnus A are consistent. 
The flux of Hydra A according to the \citet{Baars1977} flux scaling would be 1426\,$\pm$\,89\,Jy at 29.9\,MHz, in agreement with the values found in \citet{Finlay1973}. 
Both papers 4 and 5 used the GLEAM survey \citep{Hurley-Walker2017} to find their HII regions. This survey has calibrated all their data with the flux scales found in \citet{Baars1977}. We conclude that the different flux scalings for calibrators used in the various papers are consistent, therefore no scaling is needed.


\section{Building a low--frequency catalog}
\label{sec:catalog}

In this chapter we discuss what computations have been done with the data from papers 1--5 to compile them into one coherent low--frequency catalog, rescaled to 74\,MHz. The complete low-frequency catalog can be found at the end of this paper, in Table \ref{tab:catalog}.

 \subsection{Additions and updates to published data}
\label{sec:building}

\subsubsection{Brightness Temperatures}
Paper 1 has not published errors on the observed brightness temperature, T$_{\rm min}$ in their table 2. In their discussion they conclude that the absolute calibration introduces an uncertainty of roughly 2000 K, which is one part of the total error. The second part is based on the r.m.s. and is thought to be $\sim 900\,- \,4400$~K depending on the location on the sky. The weighted mean of these errors is calculated for both HII regions.

We use the electron and foreground temperatures to calculate a brightness temperature for the HII regions in paper 2. Because of an incomplete discussion on the uncertainties for their measurements we choose to use an error of 10$^4$~K on each brightness temperature. This error translates to roughly an 18 to 40 percent uncertainty of the observed temperatures. This spread is similar to the one in the uncertainties for the rest of the observed temperatures in the catalog.

Published data in paper 4 consist of integrated surface brightnesses in Jy. To compare these to the rest of the catalog data, a mean surface brightness per beam was calculated for each HII region. We did this by using the size of the HII region and calculating the number of beams covering the HII region, which gives the surface brightness in Jy/beam. Using the Rayleigh-Jeans law we calculate the brightness temperature at 88 MHz before rescaling to 74 MHz.

\subsubsection{Distances}
The distance errors in paper 1 were found in \citet{Georgelin1970}, one of multiple papers mentioned by the authors.

Paper 2  provided their distances without any uncertainties, which can be found in \citet{Georgelin1970} and \citet{Humphreys1978}. From these papers we have been able to reconstruct the error for each HII region distance, except for one, for which we have assumed an uncertainty of roughly 10\% of the distance. The mean relative distance uncertainty in the HII region sample for this is 15\%.

Paper 3 provides distances without uncertainties, but they reference \citet{Paladini2003} for their HII region distances. We used the radial velocities from this referenced paper, together with their discussion on kinematic distances, to calculate the proper distance errors and to check all the values in the catalog entries for paper 3. For one HII region the quoted distance was different from our own investigation and we have replaced their value with our new one.

\begin{figure}
  \resizebox{\hsize}{!}{\includegraphics{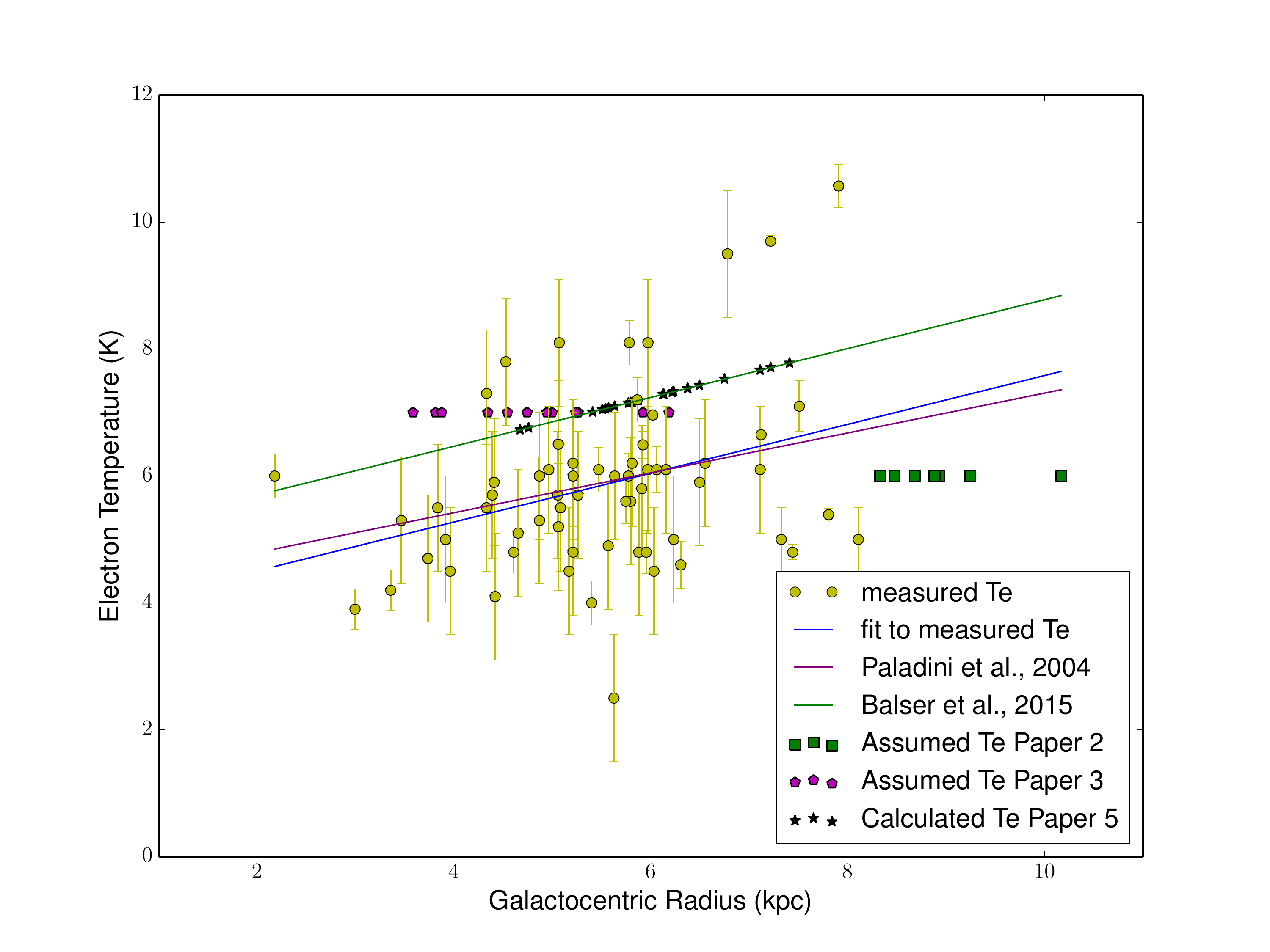}}
  \caption{Plot showing the electron temperature as a function of the galactocentric radius. The lines in this graph are a fit to the sample of measured electron temperatures in our catalog in blue; the gradient determined by \citet{Balser2015} in red; the gradient determined by \citet{Paladini2004} in green.}
  \label{fig:electrontemperatures}
\end{figure}

\subsubsection{Electron temperatures and the electron temperature gradient with Galactocentric radius}
\label{sec:electrontemp}

For 59 HII regions out of our current sample of 124, the electron temperatures have been determined through measurements of hydrogen radio recombination lines and continuum. However, for the other 65 HII regions, no measurements were available and various values for their electron temperature have been adopted: $T_{\rm e} = 10^4$~K in paper 4, $T_{\rm e} = 7000$~K in paper 3, or following the gradient with Galactocentric radius as given by \citet{Balser2015} in paper 5. We investigate whether the assumed values for $T_{e}$ need to be adapted for consistency, in our catalog. Various values of electron temperature gradients have been published \citep[e.g.,][]{Paladini2004, Alves2012, Balser2015}, which are not always in agreement with each other. However, \citet{Balser2015} concluded that there is azimuthal variation of electron temperatures of HII regions in the Milky Way as well. Therefore, gradients of electron temperature with Galactocentric radius are expected to vary for different spatial distributions of used sources.
Due to the absence of well-defined electron temperatures in paper 1, we have decided to update these for the HII regions in this set. We have used values from a more recent survey by \citet{Azcarate1990}.
Paper 2 does not discuss the origin or uncertainty of their electron temperature of 6000 K; based on the electron temperature values for the other catalog entries we have assumed an uncertainty of 1000 K.
For a sample of 19 HII regions, paper 4 does not provide electron temperatures. We have assumed an electron temperature of 7000 K $\pm$~2000 K.
All measured and assumed electron temperatures in our catalog are plotted in Fig.~\ref{fig:electrontemperatures}. It shows that the assumed values for electron temperatures are in the same temperature range as the measured temperatures. Furthermore, there is such a large variation in measured temperature between individual HII regions, that we do not think that adapting the assumed temperatures to any of the published gradients is useful. Hence, we decided to keep the assumed electron temperature values as published in their original catalogs. 
However, we tested what difference in emissivity would result from adapting the nearby (foreground) HII regions to their values as predicted by the gradient given in \citet{Paladini2004}. The emissivities as calculated with these new electron temperatures only changed by a small fraction of their error. This reinforces the conclusion that there is no value in adapting the assumed electron temperatures in this work.

\subsection{Rescaling to 74\,MHz}
\label{sec:rescale}
To be able to compare the different datasets to each other we have rescaled all to 74 MHz. This frequency was chosen when we compiled papers 1-3 into a catalog, as most of the data were already at 74 MHz. The 88 MHz data from papers 4 and 5 were not added until later. 
To rescale brightness temperatures to a new frequency we have chosen a temperature spectral index of $-2.7$. This value is also used in papers 1, 3 and 4. Papers 2 and 5 use different values of $-2.55$ and $-2.3$ respectively. Other discussions of the temperature spectral index (specifically \citet{Strong2011} and references therein) provide spectral indices ranging from $-3.0$ to $-2.3$, for frequencies ranging between 80\,MHz and 22\,GHz. We consider $-2.7$ a reasonable value since all but one of the emissivities stay within one sigma of each other when using any spectral index between $-2.3$ and $-2.7$. The outlier here still lies well within 3 sigma when using spectral index $-2.3$ instead of $-2.7$.

The temperature spectral index $\beta$ is defined as $T \propto \nu^{- \beta}$,  the temperature at the new frequency 
\begin{eqnarray}
\frac{T_{\rm 74}}{T_{\rm ori}}=\left( \frac{74 \mbox{ MHz}}{\nu_{\rm ori}}\right)^{-\beta},
\label{eq:rescale}
\end{eqnarray}

where $T_{\rm ori}$ is the foreground or background brightness temperature at the original frequency $\nu_{\rm ori}$, $T_{74}$ is the foreground or background brightness temperature at 74\,MHz. The new emissivity is defined 
as 
\begin{eqnarray}
\epsilon_{74} = \frac{T_{\rm 74}}{D},
\end{eqnarray}
where $D$ is either $D_{\rm F}$ or $D_{\rm B}$.


\section{Analysis of Observed Emissivities}
\label{sec:analysis}

\begin{figure}
  \resizebox{\hsize}{!}{\includegraphics{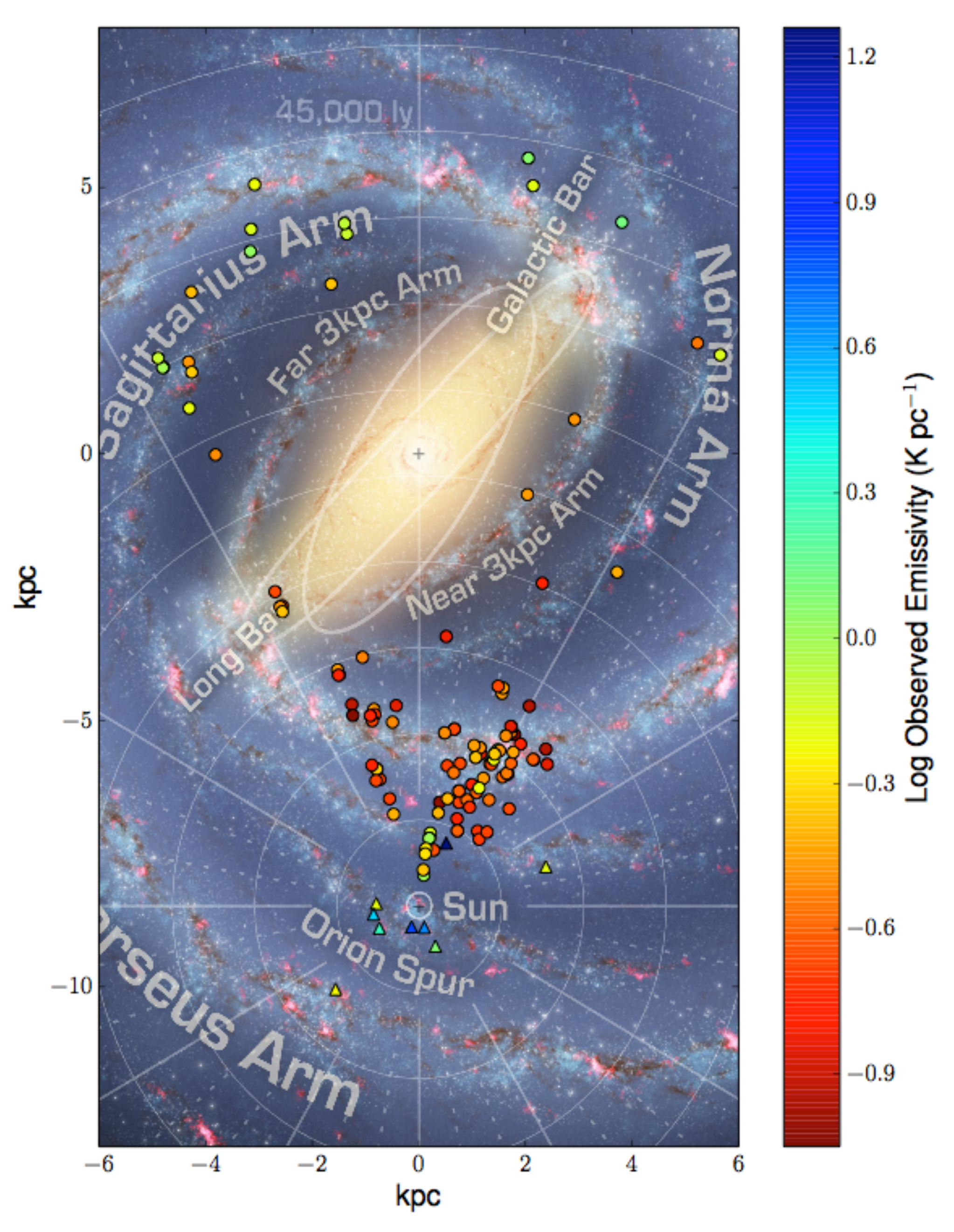}}
  \caption{Plot showing the observed emissivities as a function of the HII region location in the Milky Way, plotted in a logarithmic color scale. HII regions with foreground emissivities ($\epsilon_{\rm F}$) are indicated with a triangle, background emissivities ($\epsilon_{\rm B}$) with circles. Background image: NASA/JPL-Caltech/ESO/R. Hurt.}
  \label{fig:MWEmObs}
\end{figure}

\begin{figure*}
  \resizebox{1.0\textwidth}{!}{\includegraphics{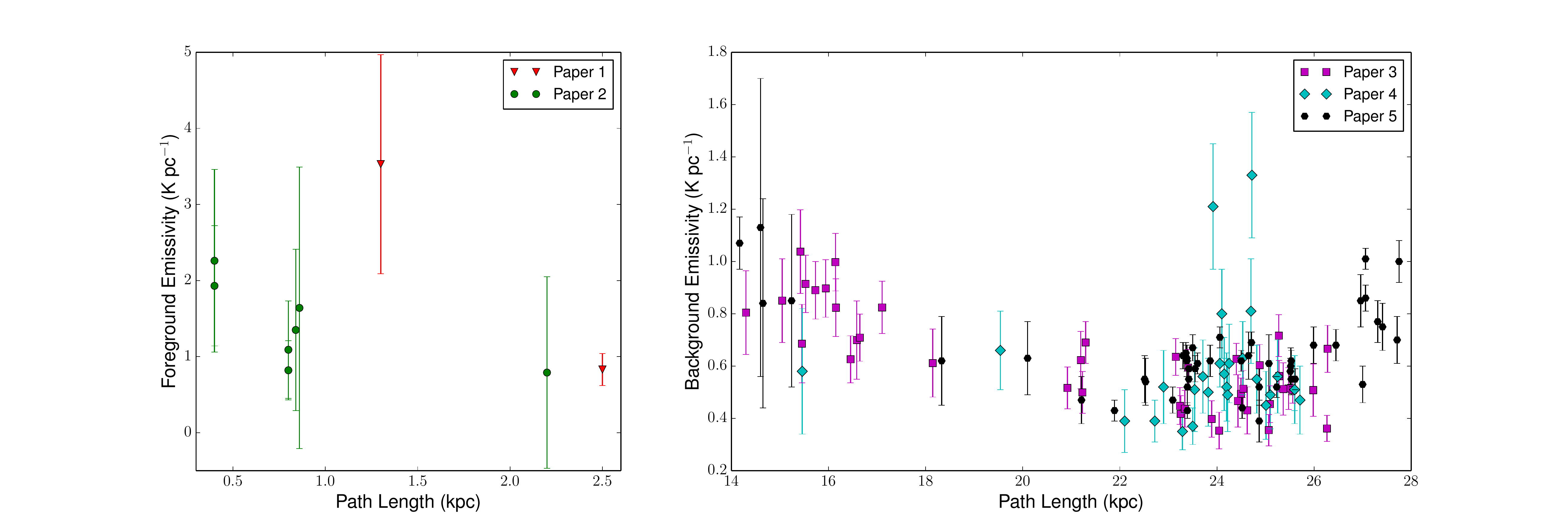}}
  \caption{On the left we show the observed foreground emissivities ($\epsilon_{\rm F}$) as a function of the path length for papers 1 and 2. On the right, the same, but for the observed background emissivities ($\epsilon_{\rm B}$) from papers 3 through 5. In Sect.~\ref{sec:analysis} we discuss these measurements in detail.}
  \label{fig:datafgbg}
\end{figure*}

\begin{figure}
  \resizebox{0.5\textwidth}{!}{\includegraphics{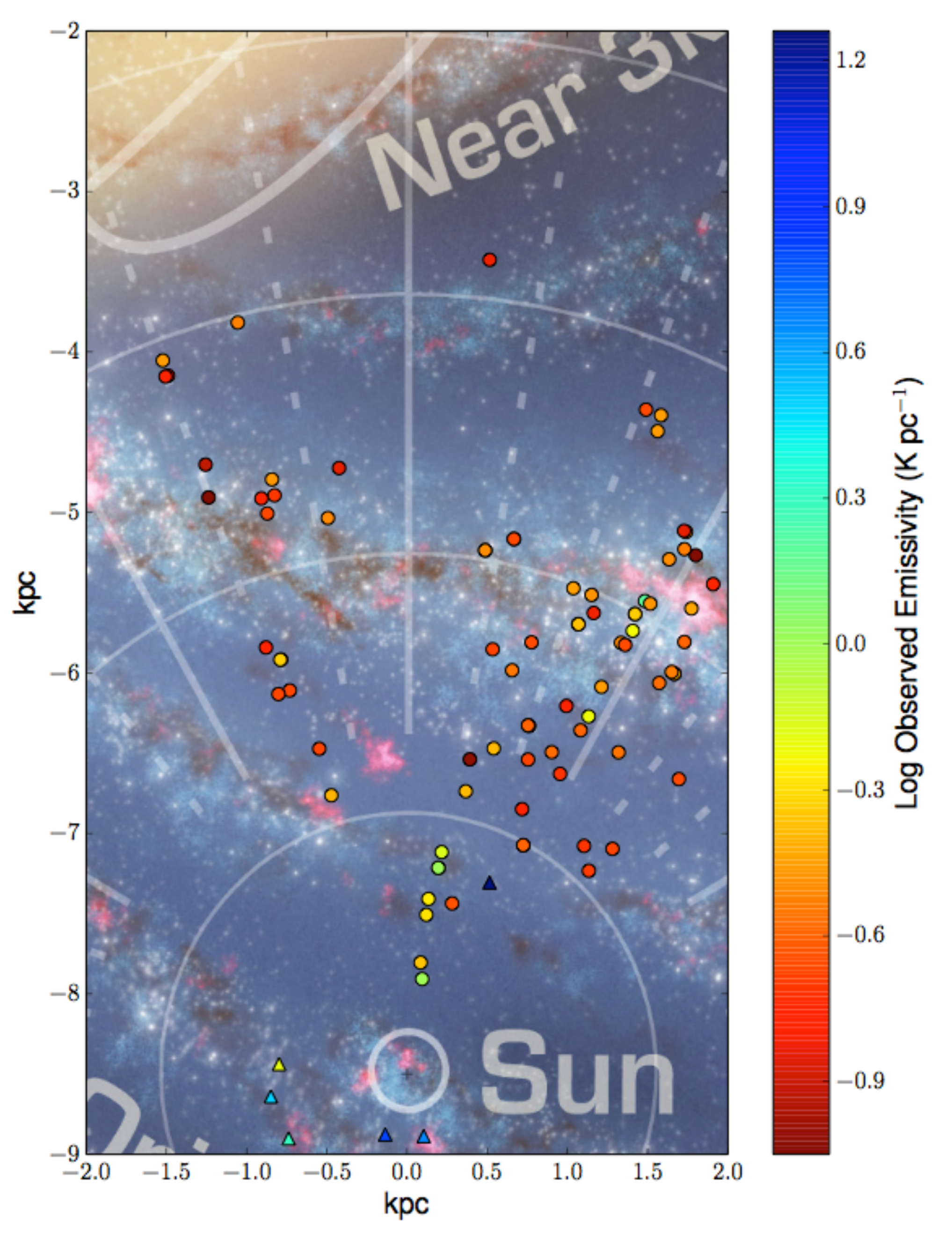}}
  \caption{A zoom-in of the Milky Way plot in Fig.~\ref{fig:MWEmObs}. The dust band moving from coordinates [$-$2,$-$5] kpc and [+2,$-$6] kpc is the Scutum-Centaurus spiral arm.}
  \label{fig:EmissObsYellowFeat}
\end{figure}

Figure~\ref{fig:MWEmObs} shows the emissivities, $\epsilon$~from the catalog overlaid on an artist's concept of the Milky Way plane. Here the triangles indicate the foreground emissivities which show mostly high values of log $\epsilon >$ 0.6. These foreground emissivities are also shown in the left hand plot of Fig.~\ref{fig:datafgbg} and are traced with HII regions rather close to the Sun ($D < 2.5$ kpc, Table~\ref{tab:catalog}). We will attempt to use a local enhancement to explain these higher emissivity values, which is discussed in further detail in Sect.~\ref{sec:localenhancement}. 
The background emissivities in Figs.~\ref{fig:MWEmObs} and~\ref{fig:datafgbg} show four noticeable features. Firstly, the general trend for HII regions behind the Galactic center (from our point of view), is to have mid range emissivities (log $\epsilon \approx -0.3$~to 0.0). These HII regions have path lengths between 14 and 17 kpc. Secondly, HII regions in front of the Galactic center (path lengths 22 -- 26 kpc) feature mostly in low range emissivities (log $\epsilon < -0.3$).
Thirdly, six HII regions on the LOS in the direction (within a few degrees) of the Galactic center do not show the lower emissivity values of the surrounding HII regions, but have values of log $\epsilon \approx$~0.0 to 0.3, indicating an increase in the emissivities along those LOS. This is also shown in the zoom in Fig.~\ref{fig:EmissObsYellowFeat} (the 6 yellow/green circles close to the Sun). These HII regions have path lengths between 27 and 28 kpc. This might mean that there is a region with higher emissivity behind this group of HII regions but in front of any path lengths that do not show this elevated emissivity. We speculate that this could be caused by a localized feature of enhanced synchrotron emission in the Scutum-Centaurus arm. Fourthly, in the right-hand plot of Fig.~\ref{fig:datafgbg}, two data points from paper 4 can be seen with slightly elevated emissivities. Considering the spatial location of these two HII regions, we conclude no deviating location: they are located within the 3D distribution of all HII regions. Therefore, it is deemed unlikely that their deviating measured values are real and caused by a local enhanced emissivity. Instead, it is suspected that the measurements may be overestimates of the real emissivities. These two HII regions in particular have an estimated surface comparable to the beam size. Therefore, possible contamination by background synchrotron emission cannot be excluded. Discarding these points would not make a significant change in the model fits.


\section{Emissivity modeling} 
\label{sec:modeling}

In a two-step attempt to model the catalog emissivities we use simple axisymmetric emissivity models and detailed GMF-based emissivity models to describe the catalog emissivities.

\subsection{Models}
\label{sec:models}
A number of different models, as used in papers 3 and 5, are employed to simulate synchrotron emissivity and optimize their parameters by comparison to the observed emissivities, using least-squares minimization. 

\begin{enumerate}
 \item constant emissivity 
 \item Gaussian decrease in synchrotron emissivity with galactocentric radius 
 \item exponential decrease in synchrotron emissivity with galactocentric radius
 \item three zone model (TZM), consisting of concentric rings centered on the Galactic center, with varying emissivities.
 \end{enumerate}
 
 The parameters for each model can be found in Table~\ref{tab:modelingequations}. The TZM is different from the others: its three emissivity values are optimised by the code, but the radii that divide these zones are fixed before the final optimization. The radii used here are chosen by running the existing optimization code with an extensive combination of radii. The R1-R2-combination that provides the best reduced Chi-squared value is chosen. The best-fit results are shown in Fig.~\ref{fig:fitfgbg}, and are discussed in Sect.~\ref{sec:resultssimple}.
 
  \begin{table}
\caption{Emissivity modeling equations}              
\label{tab:modelingequations}      
\centering                                      
\begin{tabular}{l|l }          
\hline\hline                        
{\bf model name} &{\bf equation}\\    
\hline                                   

Constant & $ \epsilon $= c\\
\hline
Gaussian & $\epsilon =  \alpha_1 \times e^{\frac{-R_{\rm gal}^2}{2 \times \beta_1^2}}$\\
\hline
Exponent & $\epsilon = \alpha_2 \times e^{-\beta_2 \times R_{\rm gal}}$ \\
\hline
Three Zone& zone1 (R$_{\rm gal}$ $<$ R$_1$) $\epsilon = \epsilon_1$\\
	&  zone2 (R$_1$ $<$ R$_{\rm gal}$ $<$ R$_2$) $\epsilon = \epsilon_2$\\
	&  zone3 (R$_{\rm gal}$ $>$ R$_2$) $\epsilon = \epsilon_3$  \\
\hline                                             
\end{tabular}
\end{table}
 
In addition to these four models we simulate Galactic emissivities using two different GMF models. Model J13b is based on the "J10" model from \citet{Jaffe2010}, but the current parameters for the model come from \citet{PlanckXLII} where it was called "Jaffe13b". Model JF12 is presented in \citet{Jansson2012}. 
Both of these models are created using total intensity and polarized synchrotron emission maps and extragalactic rotation measures. Model J13b is a 4 armed logarithmic spiral model, with a molecular ring. 
Model JF12 has 8 logarithmic spiral arm sections and a halo. Both models have a combination of coherent, random and ordered magnetic fields. The specific field strengths of all the components can be found in the references above. In general higher synchrotron intensity is observed from a region with a magnetic field direction perpendicular to our LOS.
In addition a model of a constant CR density is assumed throughout the Galaxy. 
The results for these simulations are discussed in Sect.~\ref{sec:resultscomplex}.

\subsection{Enhancement in synchrotron emissivity in the Solar neighborhood}
\label{sec:localenhancement}

A local emissivity enhancement has been used before to explain enhanced synchrotron emissivity and is discussed in various papers \citep[e.g.][]{Sun2008,Fleishman1995,Wolleben2004}. To explore any significant presence of an elevation in the foreground emissivities, we compare our catalog emissivities to a local emissivity as proposed by \citet{Beuermann1985} using a spectral index of $-2.7$. \citet{Beuermann1985} use 408 MHz data \citep{Haslam1982} to model the radio structure of our Galaxy and find a mean local emissivity of 0.011\,$\pm$\,0.003 K pc$^{-1}$, equal to 1.10\,$\pm$\,0.30 K pc$^{-1}$ at 74 MHz. This value is consistent within the error with the mean value we find for our foreground emissivities with distances within 1 kpc, which is 1.52 $\pm$~0.99 K pc$^{-1}$ at 74 MHz. This substantiates the need to model a local enhancement.

We follow the reasoning of \citet{Sun2008} and model an enhancement within 600 pc around the Sun. Exploring different enhancement values, we find that 1.2~K\,pc$^{-1}$ would provide a good fit, based on the reduced Chi-squared values (see Table~\ref{tab:modelingresults} and Sect.~\ref{sec:resultssimple}).  
As explanation for a local enhancement \citet{Sun2008} propose that there is either an enhanced local CR density or an increase of the local turbulent magnetic field, but they have not been able to distinguish between them.


\section{Modeling results}
\label{sec:results}
In this section we discuss the results of the modeling effort with the simple axisymmetric emissivity models and the more complex GMF-based emissivity models.
\subsection{Simple axisymmetric emissivity models}
\label{sec:resultssimple}

\begin{figure*}
  \resizebox{1.0\textwidth}{!}{\includegraphics{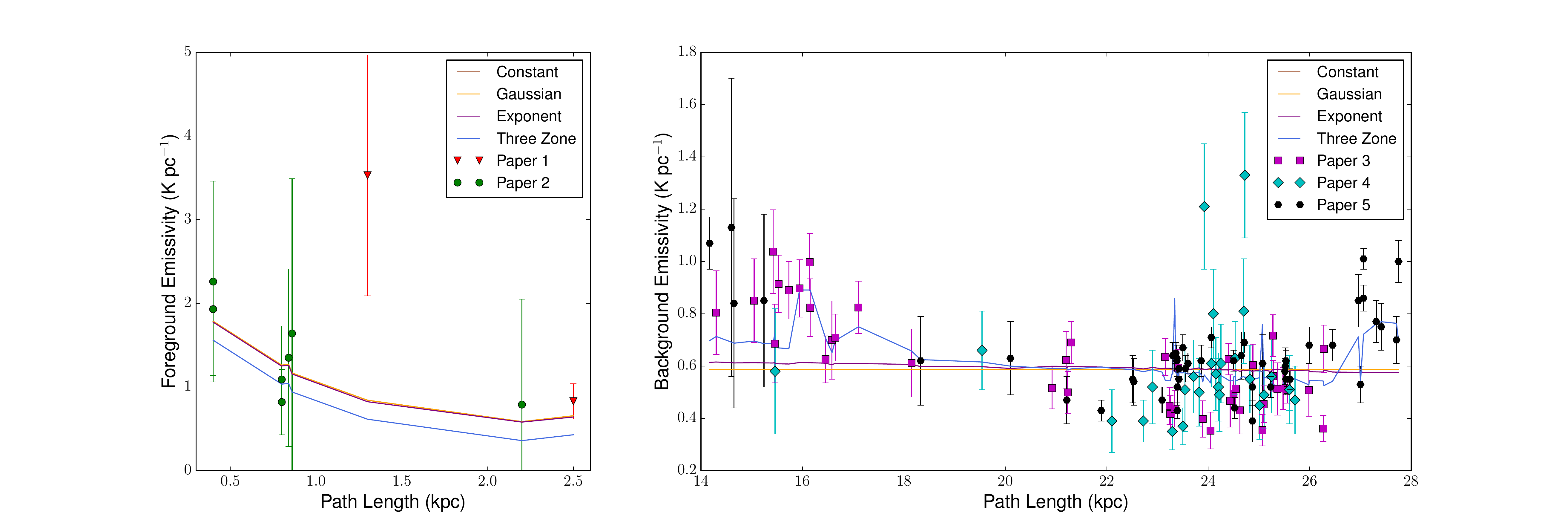}}
  \caption{On the left we show the observed foreground emissivities ($\epsilon_{\rm F}$) as a function of the path length for papers 1 and 2. On the right, the same, but for the observed background emissivities ($\epsilon_{\rm B}$) from papers 3 through 5. In both plots the simulated emissivities for the simple models are shown. In Sect.~\ref{sec:results} we discuss the results of the modeling in detail.}
  \label{fig:fitfgbg}
\end{figure*}

\begin{table*}
\caption{Emissivity modeling parameters}              
\label{tab:modelingresults}      
\centering                                      
\begin{tabular}{l|llll}          
\hline\hline                        
{\bf model name} &{\bf parameter 1 }& {\bf parameter 2} & {\bf parameter 3} &{\bf $\bar{\chi}^2$ }\\    
\hline                                   
Constant & c =  0.59\,$\pm$\,0.01 & & & 4.35 \\
\hline
Gaussian & $\alpha_1$= 0.59\,$\pm$\,0.1  & $\beta_1$= 1.45e4\,$\pm$\,2.6e9 & & 4.39 \\
\hline
Exponent & $\alpha_2$= 0.51\,$\pm$\,0.14  & $\beta_2$= $-0.01$\,$\pm$\,0.03 & & 4.38 \\
\hline
Three Zone & $\epsilon_1$ = 2.94\,$\pm$\,0.51  & $\epsilon_2$= 0.36\,$\pm$\,0.06 & $\epsilon_3$= 0.89\,$\pm$\,0.09 & 3.46 \\
& R$_1$ =  1.6$^{+2.4}_{-0.3}$ & R$_2$ = 11.0$^{+2.0}_{-1.7}$ &  &  \\
\hline                                             
\end{tabular}
\end{table*}

We compare the emissivities calculated in our models with the observed emissivities from our catalog. In Table~\ref{tab:modelingresults} we present the resulting best-fit parameters. Considering the reduced Chi-squared values from this table we can expect that the constant, Gaussian and exponential model results hardly deviate from each other. The TZM shows a better fit in general, which is not unexpected as a more complex model is likely to fit better to complex data. The combination of modeled TZM emissivity values $\epsilon_1$ and $\epsilon_2$ is consistent with the assumption that the CR density follows the source density, while the higher outer emissivity $\epsilon_3$ does not follow this naive assumption. However, this model is still too simple to have much confidence in its results. These four models were fitted to both foreground and background simultaneously, the shown fits are constrained by all the data. For clarity we have split the foreground and background emissivities, including fits, in two plots in Fig.~\ref{fig:fitfgbg}. 
For foreground emissivities the models follow the trend discussed in Sect.~\ref{sec:analysis} well. All fits are consistent with the data, within the error. 
The constant, Gaussian and exponential models show roughly the same fit and are plotted on top of each other. The TZM falls slightly below the other models, but still within two sigma from each data point. There is no significant difference between the four models, and there is no model significantly better than the rest. 
For the background emissivities, the path length spread of 14 to 28 kpc gives rise to more differences between the four models. Due to the high value of the $\beta_1$-parameter that determines the width of the bell and its even higher uncertainty, the Gaussian model is barely distinguishable from the constant model. The exponential model shows a negative slope when plotting emissivity values as a function of path length, and at shorter path lengths (from our point of view behind the Galactic center and close to the Galactic edge - see Sect.~\ref{sec:resultscomplex}) shows elevated emissivity values. Specifically regarding the GMF modeling in the next section, this is an interesting result. The exponential model does not show significant difference with respect to the constant and Gaussian models and does not fit to the observational data. The TZM is better at reproducing emissivity values compared to the other models. Nevertheless, we have to conclude that the Galactic structure cannot be reproduced by these simple axisymmetric emissivity models.

\subsection{Galactic magnetic field models}
\label{sec:resultscomplex}

Focussing on more astrophysically founded models, we use existing magnetic field models J13b and JF12. We use the software package {\sc Hammurabi} \citep{Waelkens2009} to simulate synchrotron emissivities from the GMF and (non-astrophysical) constant CR  density model.

As none of these magnetic field models take a local enhancement into account we only consider background emissivities in our analysis. The simulated data are not normalised to display emissivities at 74 MHz because the underlying magnetic field models have been based on synchrotron measurements at 22 and 94 GHz.  Due to unknown variations in the synchrotron spectral index across the Milky Way, it is difficult to reliably extrapolate to 74 MHz. Therefore, in this paper we treat the absolute flux level of the simulated data as unknown at 74 MHz.
Assuming a constant spectral index for all sight lines, we can compare the general distribution of observed and simulated emissivities as a function of HII region position. We attempt an ad hoc normalisation to the observational data by a least-squares minimizing method to be able to compare the different models and discern possible trends. It is difficult to determine if such a normalisation is correct. 
To test multiple scenarios we have performed two different normalisations. One scenario assumes that the data with long path lengths fit the model (Fig.~\ref{fig:longhamj10+jf12}), so that the simulated data sets are normalised with respect to the long path lengths ($>$ 18~Kpc). The second scenario assumes that the data with short path lengths fit the model (Fig.~\ref{fig:shorthamj10+jf12}), so that the simulated data sets are normalised with respect to the short path lengths ($<$ 18~Kpc). For convenience all data sets have been binned. Although the correctness of the normalisations cannot be determined, different trends caused by the different magnetic field models can be studied. 

\begin{figure}
  \resizebox{\hsize}{!}{\includegraphics{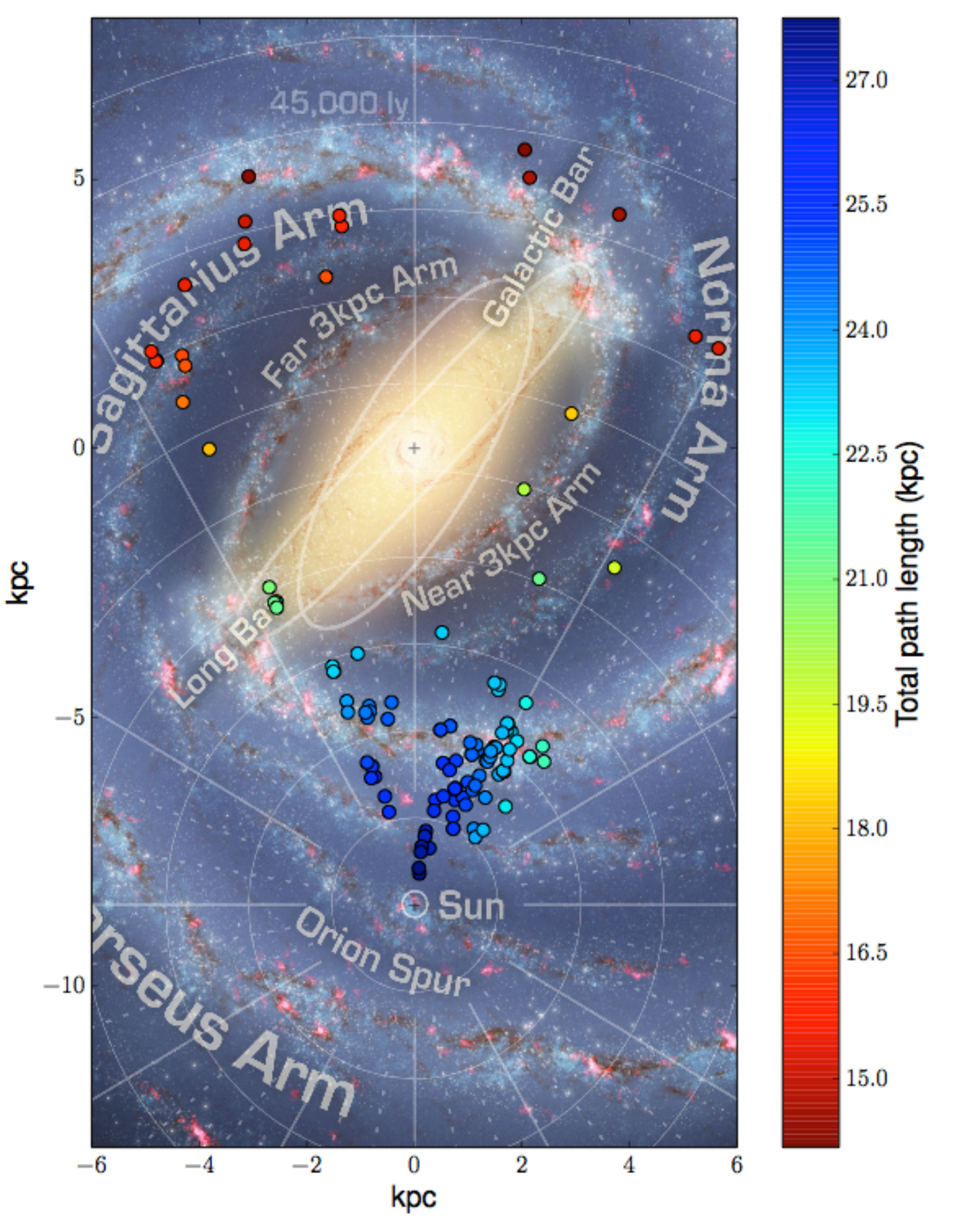}}
  \caption{Plot of the HII regions in their assumed location in the Milky Way, with a color scheme based on their LOS path lengths. HII regions with background data only.}
  \label{fig:pathlength}
\end{figure}

\begin{figure}
  \resizebox{\hsize}{!}{\includegraphics{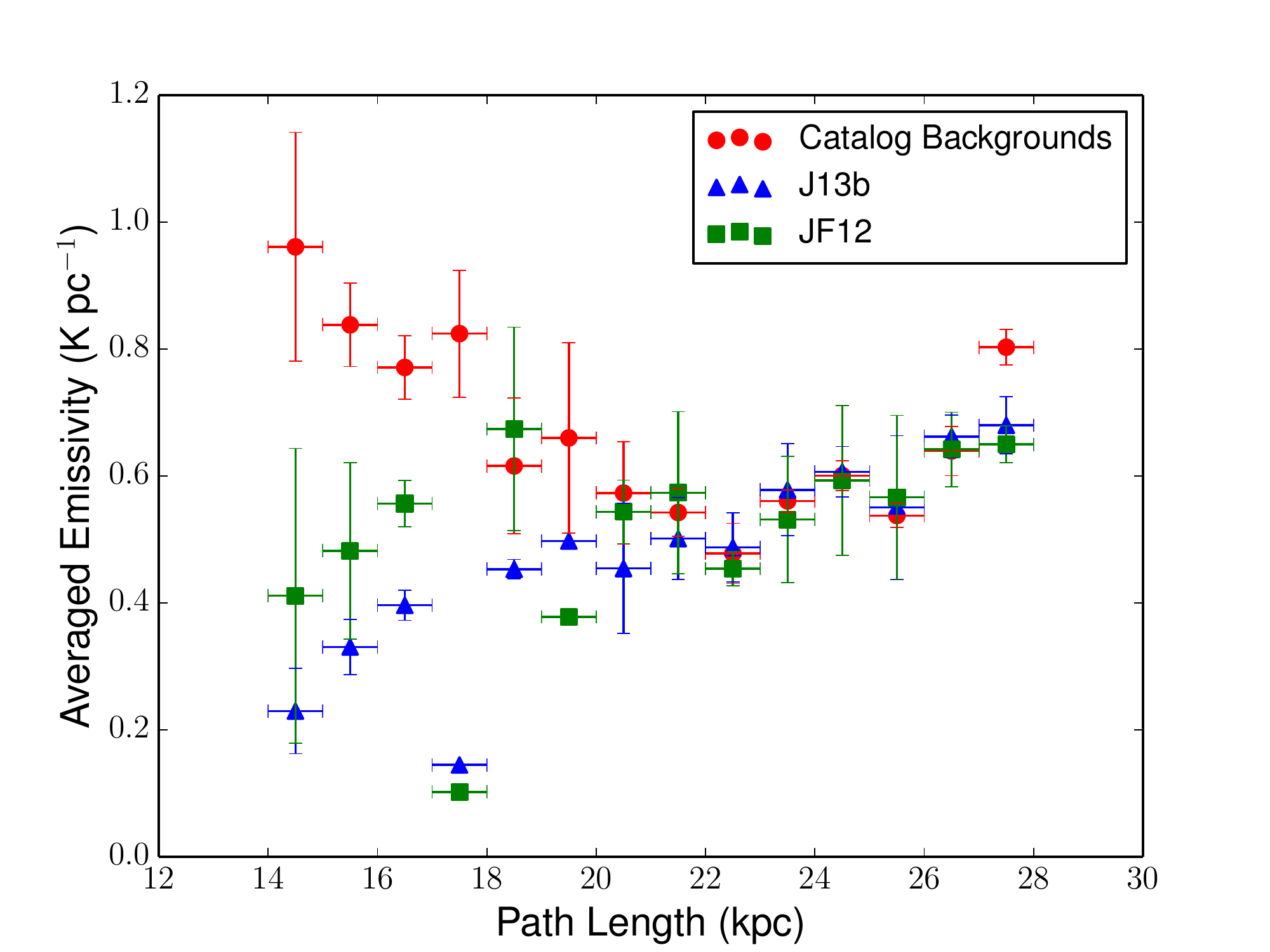}}
  \caption{Plot showing the simulated data from the GMF-based emissivity models with the observed emissivities from the catalog. All data are averaged over bins of 1~Kpc width, as indicated by the horizontal error bars. For the model data, the vertical error bars indicate the variance in the binned data. A lack of vertical error bar indicates the presence of only one data point in the bin. Shown is the synchrotron emissivity (y-axis) as a function of the path length for each HII region (x-axis). The simulated data shown here is normalised with respect to the longer path lengths ($>$ 18~Kpc). }
  \label{fig:longhamj10+jf12}
\end{figure}

\begin{figure}
  \resizebox{\hsize}{!}{\includegraphics{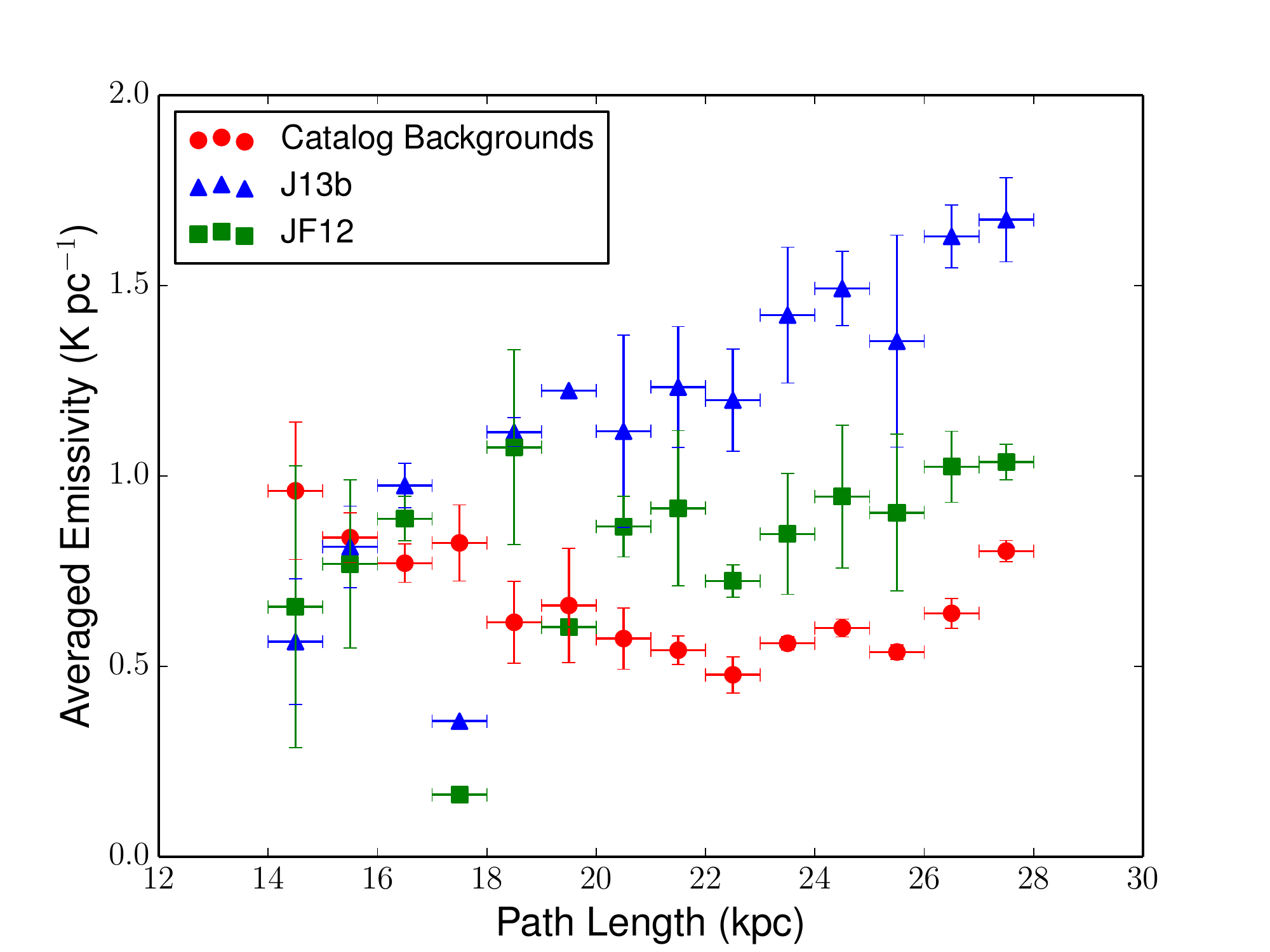}}
  \caption{Same as in Fig.~\ref{fig:longhamj10+jf12}, but the simulated data shown here is normalised with respect to the shorter path lengths ($<$ 18~Kpc).}
  \label{fig:shorthamj10+jf12}
\end{figure}

Figure~\ref{fig:longhamj10+jf12} shows a lower emissivity in models as compared to observations for path lengths between 14 and 20 kpc. We plot the location of the HII regions considered here in Fig.~\ref{fig:pathlength}, which shows that path lengths between 14 and 18 kpc are located behind the Galactic center. This is the same subgroup of HII regions we discussed in Sect.~\ref{sec:analysis}, related to higher emissivities behind the Galactic center. 

At longer path lengths and thus closer to the observer (see Fig.~\ref{fig:pathlength}), we see that the simulated values are better able to reproduce the trend in the observed emissivities. This suggests that we would need a higher CR density or stronger magnetic field in the far outer Galaxy to explain these observations.

Figure~\ref{fig:shorthamj10+jf12} shows that an agreement of observed and simulated emissivities at the short path lengths gives rise to an overprediction of emissivity at the long path lengths. 
This can be explained with a lower magnetic field strength or CR underdensity in the central region of the Galaxy. The binned figures seem to suggest that a better fit is achieved for the long path length normalisation in Fig.~\ref{fig:longhamj10+jf12}.

\begin{figure}
  \resizebox{\hsize}{!}{\includegraphics{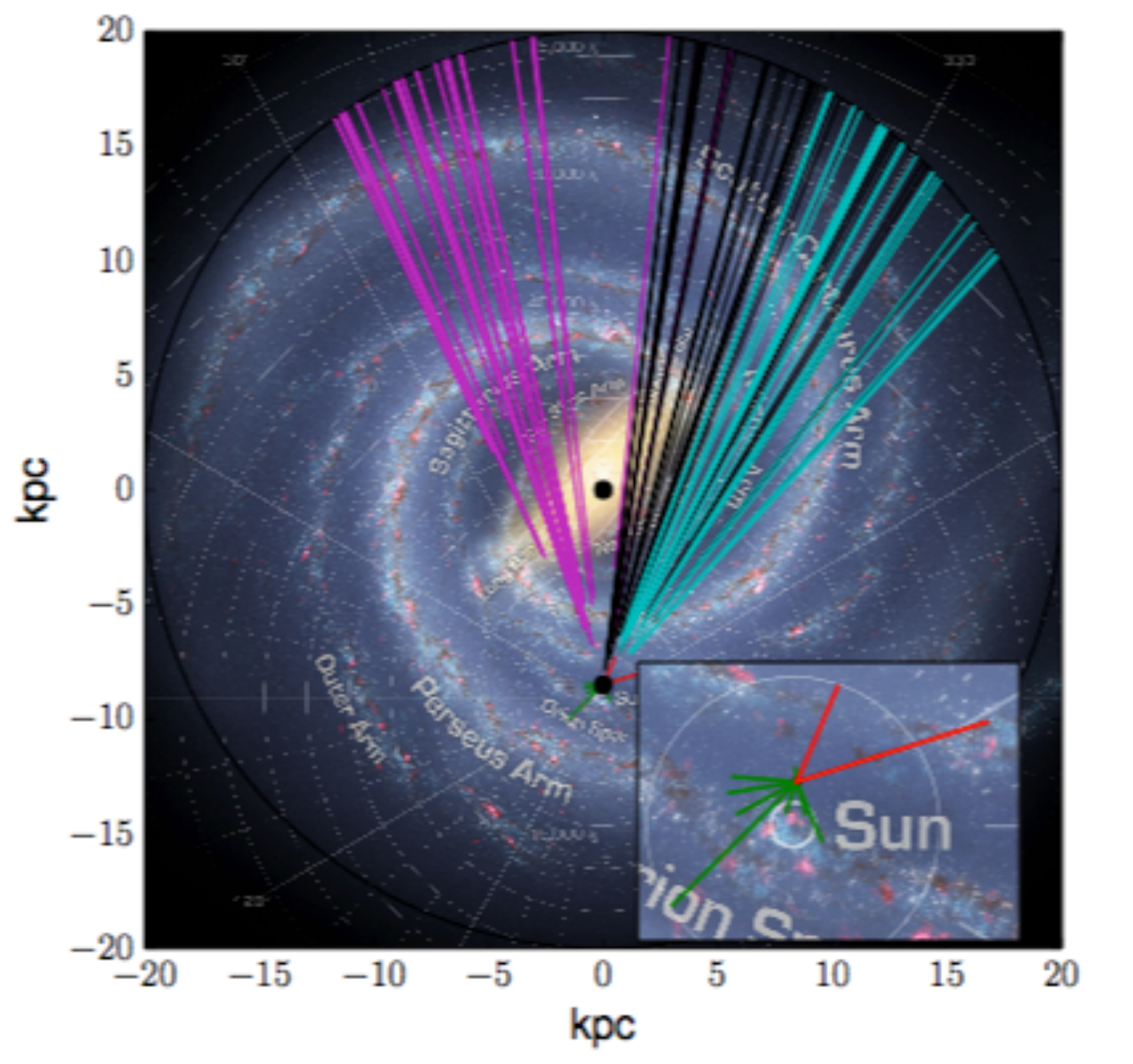}}
  \caption{Plot that shows the lines of sight for all HII regions in the catalog. Inset is a zoom-in of the foreground LOS that are close to the Sun. We show paper 1 in red, paper 2 in green, paper 3 in magenta, paper 4 in cyan and paper 5 in black.}
  \label{fig:losall}
\end{figure}


\section{Discussion}
\label{sec:discussion}

\subsection{Comparing to previous work}
\label{sec:discussionoldpapers}

One of the models put forward in paper 3 is the constant emissivity model. Their conclusion that this model is unsuitable is reinforced with the additions of multiple data sets to the catalog. 
A constant type model can be fit to the data if we split the catalog up in subsets and try a model that has several different constant emissivities in different regions, exactly like the TZM. Expanding the number of regions in this model will allow a better fit, not unexpected as one increases the degrees of freedom. However, this type of model does not have a clear physical basis from which to explain any underlying distribution of CR density or magnetic field. It will likely also not explain the scatter in emissivities that we see in our sample. 

The second model from paper 3 is one where the inner 3 kpc region in the Galactic center has a CR underdensity and the rest of the Galactic plane has a constant emissivity. This model as a whole cannot be reproduced, moreover the simple models - specifically the TZM - only produce an overdensity in this region.
 
However the modeling results in Sect.~\ref{sec:resultscomplex} do show one of two scenarios that might call for the underdensity that paper 3 has suggested.

In general terms our conclusions about the simple axisymmetric emissivity models are similar to the conclusions in paper 5. We can agree that these four models are much too simplified to explain all the features in the observations. In terms of the modeling parameters however there is little overlap between them. Considering the use of additional data to that of paper 5 these differences are expected. More complex models based on more than axisymmetry are needed.  

\subsection{This work}
\label{sec:discussionthispaper}

Analysing the results from Sect.~\ref{sec:results}, there are multiple scenarios that can play a role. In this section we try to substantiate these scenarios.

With respect to the modeled emissivities, the observed emissivities in the region behind the Galactic center appear high (Fig.~\ref{fig:longhamj10+jf12}). This can be explained by introducing either a high magnetic field strength or a high CR density in the far outer region of the Milky Way. An elevated GMF strength for this scenario would likely be located in the far branch of the Scutum-Centaurus arm, and has to cover at least an area in direction of the Galactic center between Galactic longitudes 330$^{\circ}$  and 30$^{\circ}$.  Considering that the direction of the arm in the far outer Galaxy is perpendicular to our LOS, an increased GMF strength would indeed maximally increase the emissivity from this region. It cannot be determined whether this region of enhanced magnetic field would extend only over this region, or extend along the Scutum-Centaurus arm. However, if the entire spiral arm would have an elevated field strength, this would have been apparent in the fit parameters of the J13b and JF12 models. In addition, our data does not show any evidence of enhanced magnetic field strength in the near crossing of the Scutum-Centaurus arm.
Results by \citet{Beck2015} (their Fig.~3, and related text and references therein) suggest that at least one external galaxy (IC342) shows a higher magnetic field strength at outer radii. 
However there is no indication that this is the case in the Milky Way. Figure 1 in \citet{Beck2001} shows no elevated magnetic fields in the Milky Way as a function of Galactic radius. And considering the lack of Galactic CR accelerators in this part of the Galaxy, an elevated CR density is also not expected. Overall this first scenario seems unlikely.

The lower emissivities at longer path lengths (Fig.~\ref{fig:shorthamj10+jf12}) can be achieved by either low magnetic field strength or a low CR density in the inner radii of the Galaxy.
This scenario can be connected to Galactic winds that rapidly move CRs to other parts of the Galaxy from the Galactic center. Current Galactic wind models \citep[e.g.,][]{Recchia2016,Taylor2017} do not give any indication on what level of CR displacement is achieved. This scenario can also be connected to a possible X-shaped GMF \citep{Ferriere2014}, where an additional removal of CRs is realised. It is uncertain at this time what the extent of the CR outflow will be, and if it will be sufficient. In Figure 6a in \citet{Beuermann1985} however, there is an indication that the emissivity does have a downward trend in the central radii of the Milky Way.

The third option is that none of the above scenarios are true, which is a reasonable assumption since the models are still very simple. 
Let us assume that the structure in CR density will give rise to variable spectral indices for the synchrotron emissivity. This implies that the constant spectral index value of $-2.7$ we have used is incorrect for most regions of the sky. Consequently using different spectral indices to rescale the different entries in our catalog, the current observed differences might disappear and with it the need to establish scenarios 1 and 2 above. 
In addition to a variable spectral index, we also need to consider the validity of using a radius of 20 kpc to describe the Galactic disk. 

These options will be explored in a forthcoming paper.


\section{Conclusions}
\label{sec:conclusion}
In this work we have combined existing low frequency observations of HII regions, with known distances in the Milky Way, into one catalog. We simulate the values in this catalog using different models. 
Using all these observations combined is advantageous, as the data are distributed over a larger area of the Milky Way than all the original data sets separately. This allows us to probe the emission on much wider distance scales. 
This is especially important for the simple axisymmetric emissivity models, because the parameters need to be valid for a larger distribution than before. The simplicity of these models and the lack of a physical basis is their undoing, and much more detailed models are needed.
One new feature in this work is the usage of existing GMF-based emissivity models to simulate Galactic synchrotron emission as one would observe it. 

Our comparison shows that the inconsistencies between the simulated and observed data can be explained by any of the following scenarios:
\begin{enumerate}
\item elevated CR density or magnetic field strength in the outer radii of the Milky Way
\item lower CR density or magnetic field strength in the inner radii of the Milky Way
\item a variable spectral index that changes the morphology of the current emission level distribution.
\end{enumerate}

Due to a considerable body of work on both the magnetic field strength and the likely CR density in the outer region of the Milky Way, the first scenario seems unlikely. The second and third scenarios however appear promising and will be discussed in upcoming paper(s), since theories of Galactic winds and X-shaped magnetic fields might provide a big enough outflow of CRs to cause a paucity in the inner region of the Galaxy. And considering the unlikely assumption of a constant spectral index in the entire Galaxy, new rescaled emissivities need to be calculated for the next version of our catalog. These new values can be used to revisit some of the modeling here.

\section{Future Work}
\label{sec:futurework}

As already set forth in the Discussion and Conclusion sections, a non-constant spectral index is likely to change the distribution of the synchrotron emissivities in our collection. An important future venture is to use the measured spectral-index spread along the Milky Way to redo some of the modeling presented in this paper. It will be interesting to see the likely change in scenarios that are put forth by the modeling effort now.

Another interesting change in the current effort would be the Galactic boundary, which is currently set to 20~kpc. This value is the assumed boundary of the region emitting the bulk of the Galactic synchrotron emission. The actual value of the Milky Way boundary is unknown, though it is possible that CRs and magnetic fields will extend far beyond those 20~kpc \citep{Levine2006}. It would be an interesting exercise to quantify the changes with a different Galactic radius.

Considering the current conclusions and the suggested scenarios, a worthwhile pursuit will be the comparison of more GMF-based emissivity models. If a constant CR density model (like the one in this paper) is used, the simulated synchrotron emissivity from many different GMF-based emissivity models can be compared to both the existing catalog and to each other. The different details for each GMF and their influence on the simulated emission can be discussed. Alternatively more realistic CR distributions (e.g., GALPROP \citep{Strong2009}, DRAGON \citep{Maccione2011}, PICARD \citep{Kissmann2014}) can also provide interesting results when used in our simulations.

New emissivity values for HII regions with known distances can be added to the existing catalog, specifically from observations with the LOFAR instrument \citep{VanHaarlem2013}. We hope to add HII regions from a signicant part of the Milky Way plane to the catalog, e.g. using results from the ongoing Low-Band Array surveys with LOFAR. The expanded catalog can be used for extended modeling of the Galactic synchrotron emissivity. 

\section*{Acknowledgements}
IP would like to acknowledge funding from the Netherlands Research School for Astronomy (NOVA). Cameron van Eck is thanked for his help and advice, specifically regarding modeling and plotting. This research made extensive use of NumPy \citep{vanderWalt2011}; IPython \citep{Perez2007} and matplotlib \citep{Hunter2007}.

\bibliography{P1}

\longtab{5}{
\onecolumn{
\begin{longtable}{*{7}{c}}

\caption{\label{tab:catalog} The complete catalog of HII regions detected in absorption. Column (1) contains the Galactic coordinates, Column (2) is the distance from the Sun to the HII region, Column (3) is the sky brightness temperature derived from the measured intensity at the observing frequency, Column (4) indicates the synchrotron brightness temperature of the column in front of the HII region (first 9 entries), and the synchrotron brightness temperature of the column behind the HII region (entries 10 and onward), Column (5) is the electron temperature of the HII region, and Column (6) is the emissivity of the column in front of the HII regions (first 9 entries) and behind the HII region (entries 10 and onward), Column (7) indicates the origin paper of the data with [1] \citet{Jones1974}, [2] \citet{Roger1999}, [3] \citet{Nord2006}, [4] \citet{Hindson2016} and [5] \citet{Su2016}.}\\
\hline
{\bf $\ell$~\textpm~b}&{\bf Distance}&{\bf T$_{\rm{obs}}$}&{\bf T$_{\rm F}$}&{\bf T$_{\rm e}$}&{\bf ${ \epsilon}_{\rm F}$}&{\bf Source}\\
&kpc&$\times10^3$~K&$\times10^3$~K&$\times10^3$~K&K pc$^{-1}$&\\
(1)&(2)&(3)&(4)&(5)&(6)&(7)\\
\hline
\endfirsthead
\caption{continued.}\\
\hline
{\bf $\ell$~\textpm~b}&{\bf Distance}&{\bf T$_{\rm{obs}}$}&{\bf T$_{\rm B}$}&{\bf T$_{\rm e}$}&{\bf $\epsilon_{\rm B}$}&{\bf Source}\\
&kpc&$\times10^3$~K&$\times10^3$~K&$\times10^3$~K&K pc$^{-1}$&\\
(1)&(2)&(3)&(4)&(5)&(6)&(7)\\
\hline
\endhead
\hline
\\
\endfoot
287.5--0.5&2.5\,$\pm$\,0.625&29.0\,$\pm$\,2.554&2.08\,$\pm$\,0.12&5.0\,$\pm$\,0.5\,\footnotemark[4]&0.83\,$\pm$\,0.21&[1]\\
336.7--1.3&1.3\,$\pm$\,0.325&58.0\,$\pm$\,3.605&4.59\,$\pm$\,1.48&5.0\,$\pm$\,0.5\,\footnotemark[4]&3.53\,$\pm$\,1.44&[1]\\
\hline

85.5--1.0&0.8\,$\pm$\,0.29&23.4\,\footnotemark[2]\,$\pm$\,10.0&0.66\,$\pm$\,0.2&6.0\,$\pm$\,1.0\,\footnotemark[3]&0.82\,$\pm$\,0.39&[2]\\
99.3+3.7&0.86\,$\pm$\,0.055&43.4\,\footnotemark[2]\,$\pm$\,10.0&1.41\,$\pm$\,1.58&6.0\,$\pm$\,1.0\,\footnotemark[3]&1.64\,$\pm$\,1.85&[2]\\
118.5+6.0&0.84\,\footnotemark[5]\,$\pm$\,0.084&36.1\,\footnotemark[2]\,$\pm$\,10.0&1.14\,$\pm$\,0.88&6.0\,$\pm$\,1.0\,\footnotemark[3]&1.35\,$\pm$\,1.06&[2]\\
134.8+0.9&2.2\,$\pm$\,0.097&52.0\,\footnotemark[2]\,$\pm$\,10.0&1.74\,$\pm$\,2.77&6.0\,$\pm$\,1.0\,\footnotemark[3]&0.79\,$\pm$\,1.26&[2]\\
160.1--12.3&0.4\,$\pm$\,0.039&29.9\,\footnotemark[2]\,$\pm$\,10.0&0.9\,$\pm$\,0.47&6.0\,$\pm$\,1.0\,\footnotemark[3]&2.26\,$\pm$\,1.2&[2]\\
195.1--12.0&0.4\,$\pm$\,0.04&26.4\,\footnotemark[2]\,$\pm$\,10.0&0.77\,$\pm$\,0.31&6.0\,$\pm$\,1.0\,\footnotemark[3]&1.93\,$\pm$\,0.79&[2]\\
202.9+2.2&0.8\,$\pm$\,0.25&29.1\,\footnotemark[2]\,$\pm$\,10.0&0.87\,$\pm$\,0.43&6.0\,$\pm$\,1.0\,\footnotemark[3]&1.09\,$\pm$\,0.64&[2]\\
\hline
{\bf $\ell$~\textpm~b}&{\bf Distance}&{\bf T$_{\rm{obs}}$}&{\bf T$_{\rm B}$}&{\bf T$_{\rm e}$}&{\bf $\epsilon_{\rm B}$}&{\bf Source}\\
&kpc&$\times10^3$~K&$\times10^3$~K&$\times10^3$~K&K pc$^{-1}$&\\
(1)&(2)&(3)&(4)&(5)&(6)&(7)\\
\hline
6.1--0.6                  &   12.7\,$\pm$\,0.09          & --7.0  \,$\pm$\,1.0\,\footnotemark[1]     &     14.0\,$\pm$\,1.8                 & 7.0\,$\pm$\,2.0&  0.89\,$\pm$\,0.11                     &[3]\\
6.2--0.6                  & 12.9\,$\pm$\,0.11          & --7.2  \,$\pm$\,1.0\,\footnotemark[1]                                                        &14.2\,$\pm$\,1.7                 &7.0\,$\pm$\,2.0&0.92\,$\pm$\,0.11                      &[3]\\
6.4--0.5                  & 3.8\,$\pm$\,0.11            & --3.6  \,$\pm$\,1.0      &       10.6\,$\pm$\,2.2                 & 7.0\,$\pm$\,2.0&  0.43\,$\pm$\,0.09                    &[3]\\
8--0.2                    &11.8\,$\pm$\,0.05       & --4.6\,$\pm$\,1.4        &      11.6\,$\pm$\,2.4                 & 7.0\,$\pm$\,2.0 &  0.70\,$\pm$\,0.15                    &[3]\\ 
8.1+0.2                  & 3.5\,$\pm$\,0.10            & --8.5  \,$\pm$\,1.8    &15.0\,$\pm$\,2.2                 & 7.0\,$\pm$\,2.0& 0.60\,$\pm$\,0.08                      &[3]\\
12.7--0.2               &4.8\,$\pm$\,0.04            & --9.4\,$\pm$\,1.2      &13.9\,$\pm$\,1.6                & 4.5\,$\pm$\,1.0 &  0.59\,$\pm$\,0.07                     &[3]\\ 
12.8--0.2              & 3.8\,$\pm$\,0.06  & --9.3\,$\pm$\,1.2     &15.3\,$\pm$\,1.6                & 6.0\,$\pm$\,1.0 &  0.63\,$\pm$\,0.06                     &[3]\\
12.8+0.4               & 13.9\,$\pm$\,0.09 & --4.5\,$\pm$\,1.2                                      &11.5\,$\pm$\,2.3                 & 7.0\,$\pm$\,2.0 & 0.81\,$\pm$\,0.16                     &[3]\\ 
12.9--0.2              & 3.7\,$\pm$\,0.06  & --6.0\,$\pm$\,1.2                                      &12.1\,$\pm$\,1.6                & 6.1\,$\pm$\,1.0 & 0.49\,$\pm$\,0.06                      &[3]\\
13.9--0                 & 13.1\,$\pm$\,0.07   & --5.8\,$\pm$\,1.3                                      &12.8\,$\pm$\,2.4                 & 7.0\,$\pm$\,2.0&  0.85\,$\pm$\,0.16                     &[3]\\
14--0.1                 &  3.6\,$\pm$\,0.06 & --7.0\,$\pm$\,1.3                                      &12.6\,$\pm$\,1.7                 & 5.5\,$\pm$\,1.0&  0.51\,$\pm$\,0.07                     &[3]\\
14.2--0.2               &  3.7\,$\pm$\,0.06 & --4.4\,$\pm$\,1.4                                     &11.4\,$\pm$\,2.4                 & 7.0\,$\pm$\,2.0&  0.47\,$\pm$\,0.10                      &[3]\\
14.4--0.1               &  12.7\,$\pm$\,0.06 & --9.0\,$\pm$\,1.4                                  &16.0\,$\pm$\,2.4                & 7.0\,$\pm$\,2.0&  1.04\,$\pm$\,0.16                      &[3]\\
15.1--0.7               &  2.1\,$\pm$\,0.09 & --7.3 $\pm$2.3                                    &13.2\,$\pm$\,2.5                & 5.9\,$\pm$\,1.0 &  0.51\,$\pm$\,0.10                     &[3]\\ 
15.2--0.6              & 1.8\,$\pm$\,0.10 & --8.0\,$\pm$\,2.3                                  &17.5\,$\pm$\,2.5                & 9.5\,$\pm$\,1.0 &  0.66\,$\pm$\,0.09                       &[3]\\ 
16.9+0.8              &  2.7\,$\pm$\,0.07 & --8.1\,$\pm$\,1.4\,\footnotemark[1]                              &14.2\,$\pm$\,1.6                & 6.1\,$\pm$\,1.0 & 	 0.56\,$\pm$\,0.06                  &[3]\\ 
17.0+0.8              &  2.5\,$\pm$\,0.07  & --7.0\,$\pm$\,1.6                            &13.1\,$\pm$\,1.9                & 6.1\,$\pm$\,1.0 & 	 0.52\,$\pm$\,0.08                  &[3]\\ 
17.0+0.9              & 2.7\,$\pm$\,0.07 & --10.0\,$\pm$\,1.6                                     &18.1\,$\pm$\,1.9                & 8.1\,$\pm$\,1.0 & 	 0.72\,$\pm$\,0.08                &[3]\\ 
18.3--0.3              & 4.0\,$\pm$\,0.05 & --4.2\,$\pm$\,1.3                                       &9.5\,$\pm$\,1.8                   & 5.3\,$\pm$\,1.0 &  0.40\,$\pm$\,0.07                    &[3]\\
18.3+1.9              & 2.8\,$\pm$\,0.07 & --5.6\,$\pm$\,1.5                                       &11.4\,$\pm$\,1.8                 & 5.8\,$\pm$\,1.0 &  0.46\,$\pm$\,0.07                      &[3]\\
18.7+2.0               & 2.5\,$\pm$\,0.07 & --6.0\,$\pm$\,1.5                                 &13.0\,$\pm$\,2.5                & 7.0\,$\pm$\,2.0&  0.51\,$\pm$\,0.10                          &[3]\\
18.9--0.5               & 4.6\,$\pm$\,0.04 & --3.8\,$\pm$\,1.2                                  &9.7\,$\pm$\,1.6                  & 5.9\,$\pm$\,1.0&  0.42\,$\pm$\,0.07                         &[3]\\
18.9--0.4               &4.7\,$\pm$\,0.04 & --9.2\,$\pm$\,1.2                                   &14.7\,$\pm$\,1.6                & 5.5\,$\pm$\,1.0&  0.63\,$\pm$\,0.07                         &[3]\\
19.0--0.0               & 4.0\,\footnotemark[6]\,$\pm$\,0.07 & --3.3\,$\pm$\,1.2              &8.5\,$\pm$\,1.6                  & 5.2\,$\pm$\,1.0& 0.35\,$\pm$\,0.07                           &[3]\\
19.1--0.3               &4.6\,$\pm$\,0.04 & --6.3\,$\pm$\,1.2            &10.4\,$\pm$\,1.6                & 4.1\,$\pm$\,1.0&  0.45\,$\pm$\,0.07                          &[3]\\
20.3--0.9               &12.3\,$\pm$\,0.05 & --3.6\,$\pm$\,1.1                &10.6\,$\pm$\,2.3                & 7.0\,$\pm$\,2.0&  0.69\,$\pm$\,0.15                          &[3]\\
22.9--0.3               & 11.1\,$\pm$\,0.04 & --5.2\,$\pm$\,1.1             &10.3\,$\pm$\,1.5                & 5.1\,$\pm$\,1.0&  0.62\,$\pm$\,0.09                          &[3]\\
23.0--0.4               &10.9\,$\pm$\,0.04 & --4.0\,$\pm$\,1.1               &11.8\,$\pm$\,1.5                 & 7.8\,$\pm$\,1.0& 0.71\,$\pm$\,0.09                          &[3]\\
24.2+0.2               &   9.3\,$\pm$\,0.05 & --4.1\,$\pm$\,1.3                   &11.1\,$\pm$\,2.4                 & 7.0\,$\pm$\,2.0& 0.61\,$\pm$\,0.13                          &[3]\\
24.4+0.1               &  6.2\,$\pm$\,0.05 & --5.1\,$\pm$\,1.3                 &10.6\,$\pm$\,1.6                 & 5.5\,$\pm$\,1.0&  0.50\,$\pm$\,0.08                         &[3]\\
24.5+0.2              & 6.5\,$\pm$\,0.05 & --6.1\,$\pm$\,1.3       &10.8\,$\pm$\,1.7                 & 4.7\,$\pm$\,1.0 &  0.51\,$\pm$\,0.08                        &[3]\\
24.7--0.2              & 10.3\,$\pm$\,0.04 & --8.4\,$\pm$\,1.3      &14.1\,$\pm$\,1.7                & 5.7\,$\pm$\,1.0 &  0.82\,$\pm$\,0.10                         &[3]\\
24.8--0.1              & 6.2\,$\pm$\,0.05 & --6.2\,$\pm$\,1.3         &13.2\,$\pm$\,2.4                 & 7.0\,$\pm$\,2.0 &  0.62\,$\pm$\,0.11                        &[3]\\
24.8+0.1              &  6.1\,$\pm$\,0.05 & --9.8\,$\pm$\,1.4      &14.7\,$\pm$\,1.7                 & 5.0\,$\pm$\,1.0 &  0.69\,$\pm$\,0.08                      &[3]\\
25.3--0.3              &11.2\,$\pm$\,0.05 & --7.6\,$\pm$\,1.5         &13.3\,$\pm$\,1.8                 & 5.7\,$\pm$\,1.0 &  0.82\,$\pm$\,0.11                     &[3]\\
25.4--0.3              & 11.2\,$\pm$\,0.05 & --8.0\,$\pm$\,1.5         &16.1\,$\pm$\,1.8                 & 8.1\,$\pm$\,1.0 &  1.00\,$\pm$\,0.11                       &[3]\\
25.4--0.2             & 11.4\,$\pm$\,0.05& --8.3\,$\pm$\,1.5         &14.3\,$\pm$\,1.8                 & 6.0\,$\pm$\,1.0 &  0.90\,$\pm$\,0.11                       &[3]\\
348.6--0.6           & 2.7\,$\pm$\,0.10 & --8.2\,$\pm$\,0.9          &13.0\,$\pm$\,1.4                 & 4.8\,$\pm$\,1.0 & 0.51\,$\pm$\,0.05                       &[3]\\
348.7--1              & 2.0\,$\pm$\,0.12 & --3.3\,$\pm$\,0.9       &9.5  \,$\pm$\,1.4                 & 6.2\,$\pm$\,1.0 & 0.36\,$\pm$\,0.05                       &[3]\\
351.5--0.5           & 3.3\,$\pm$\,0.10 & --3.2\,$\pm$\,1.0         &8.9  \,$\pm$\,1.4                 & 5.7\,$\pm$\,1.0 &  0.36\,$\pm$\,0.06                      &[3]\\
354.2--0.1           & 5.1\,$\pm$\,0.06 & --4.9\,$\pm$\,1.3        &10.2\,$\pm$\,1.6                & 5.3\,$\pm$\,1.0 &  0.44\,$\pm$\,0.07                       &[3]\\
\hline
\clearpage
339.18--0.42&3.0\,$\pm$\,0.4&--14.98\,$\pm$\,3.57&32.86\,$\pm$\,5.91&5.6\,$\pm$\,1.0&1.33\,$\pm$\,0.24&[4]\\

338.95+0.59&2.1\,$\pm$\,0.3&--1.22\,$\pm$\,0.53&13.12\,$\pm$\,3.3&7.0\,$\pm$\,2.0&0.51\,$\pm$\,0.13&[4]\\
336.52--1.50&1.8\,$\pm$\,0.3&--0.57\,$\pm$\,0.29&12.09\,$\pm$\,3.23&7.0\,$\pm$\,2.0&0.47\,$\pm$\,0.13&[4]\\
336.59--1.81&2.5\,$\pm$\,0.7&--0.04\,$\pm$\,0.19&11.24\,$\pm$\,3.21&7.0\,$\pm$\,2.0&0.45\,$\pm$\,0.13&[4]\\
335.78+0.01&2.2\,$\pm$\,0.3&--1.79\,$\pm$\,0.81&14.03\,$\pm$\,3.45&7.0\,$\pm$\,2.0&0.56\,$\pm$\,0.14&[4]\\
333.71--0.46&11.8\,$\pm$\,0.4&--3.08\,$\pm$\,2.05&8.91\,$\pm$\,3.65&2.5\,$\pm$\,1.0&0.58\,$\pm$\,0.24&[4]\\
333.61--0.09&3.0\,$\pm$\,0.5&--2.31\,$\pm$\,1.03&14.87\,$\pm$\,3.59&7.0\,$\pm$\,2.0&0.61\,$\pm$\,0.15&[4]\\
333.64--0.22&3.2\,$\pm$\,0.4&--3.06\,$\pm$\,0.81&14.79\,$\pm$\,2.05&6.2\,$\pm$\,1.0&0.61\,$\pm$\,0.09&[4]\\
333.33--0.39&2.7\,$\pm$\,0.3&--2.7\,$\pm$\,0.86&15.48\,$\pm$\,3.48&7.0\,$\pm$\,2.0&0.63\,$\pm$\,0.14&[4]\\
333.29--0.30&3.3\,$\pm$\,1.1&--11.19\,$\pm$\,2.8&29.03\,$\pm$\,5.49&7.0\,$\pm$\,2.0&1.21\,$\pm$\,0.24&[4]\\
333.20--0.10&2.4\,$\pm$\,0.3&--1.49\,$\pm$\,0.6&13.55\,$\pm$\,3.33&7.0\,$\pm$\,2.0&0.55\,$\pm$\,0.13&[4]\\
333.04+2.03&1.6\,$\pm$\,0.6&--2.01\,$\pm$\,0.8&12.94\,$\pm$\,2.05&6.1\,$\pm$\,1.0&0.51\,$\pm$\,0.08&[4]\\
333.07+0.02&2.5\,$\pm$\,1.1&--5.51\,$\pm$\,2.36&19.97\,$\pm$\,4.94&7.0\,$\pm$\,2.0&0.81\,$\pm$\,0.2&[4]\\
333.06--0.45&3.0\,$\pm$\,0.3&--0.91\,$\pm$\,0.35&12.63\,$\pm$\,3.24&7.0\,$\pm$\,2.0&0.52\,$\pm$\,0.13&[4]\\
333.01--0.62&3.1\,$\pm$\,1.1&--5.07\,$\pm$\,1.59&19.27\,$\pm$\,4.07&7.0\,$\pm$\,2.0&0.8\,$\pm$\,0.17&[4]\\
332.98+1.78&2.1\,$\pm$\,1.4&--0.71\,$\pm$\,0.39&12.3\,$\pm$\,3.25&7.0\,$\pm$\,2.0&0.49\,$\pm$\,0.13&[4]\\
331.15--0.52&4.3\,$\pm$\,0.4&--1.03\,$\pm$\,0.45&8.82\,$\pm$\,1.75&4.5\,$\pm$\,1.0&0.39\,$\pm$\,0.08&[4]\\
330.89--0.37&3.7\,$\pm$\,0.4&--0.16\,$\pm$\,0.16&8.07\,$\pm$\,1.62&4.9\,$\pm$\,1.0&0.35\,$\pm$\,0.07&[4]\\
329.36+0.12&7.3\,$\pm$\,0.1&--0.79\,$\pm$\,1.57&12.91\,$\pm$\,2.98&7.3\,$\pm$\,1.0&0.66\,$\pm$\,0.15&[4]\\

327.18--0.60&2.9\,$\pm$\,1.0&--1.39\,$\pm$\,0.36&13.39\,$\pm$\,3.24&7.0\,$\pm$\,2.0&0.56\,$\pm$\,0.14&[4]\\
326.67+0.57&2.4\,$\pm$\,0.3&--1.68\,$\pm$\,0.62&13.85\,$\pm$\,3.34&7.0\,$\pm$\,2.0&0.57\,$\pm$\,0.14&[4]\\
326.23+0.72&3.0\,$\pm$\,0.4&--0.47\,$\pm$\,0.18&8.74\,$\pm$\,1.62&5.0\,$\pm$\,1.0&0.37\,$\pm$\,0.07&[4]\\
322.19+0.57&1.8\,$\pm$\,0.9&--0.39\,$\pm$\,0.39&11.79\,$\pm$\,3.25&7.0\,$\pm$\,2.0&0.49\,$\pm$\,0.14&[4]\\
321.15--0.55&3.8\,$\pm$\,0.5&--0.83\,$\pm$\,1.25&8.51\,$\pm$\,2.55&4.5\,$\pm$\,1.0&0.39\,$\pm$\,0.12&[4]\\
318.19--0.59&1.7\,$\pm$\,0.3&--0.43\,$\pm$\,0.18&11.86\,$\pm$\,3.21&7.0\,$\pm$\,2.0&0.5\,$\pm$\,0.13&[4]\\
317.62--0.40&1.9\,$\pm$\,0.6&--0.55\,$\pm$\,1.09&12.05\,$\pm$\,3.64&7.0\,$\pm$\,2.0&0.51\,$\pm$\,0.16&[4]\\
317.33+0.26&2.5\,$\pm$\,0.3&--0.48\,$\pm$\,0.25&11.95\,$\pm$\,3.22&7.0\,$\pm$\,2.0&0.52\,$\pm$\,0.14&[4]\\
\hline
317.988--00.754&3.6\,$\pm$\,1.1&--1.33\,$\pm$\,0.28&9.47\,$\pm$\,0.74&4.6\,$\pm$\,0.37&0.43\,$\pm$\,0.04&[5]\\
322.036+00.625&3.5\,$\pm$\,3.5&--0.52\,$\pm$\,0.23&12.47\,$\pm$\,0.64&7.29\,$\pm$\,0.33&0.55\,$\pm$\,0.09&[5]\\
322.220+00.504&3.5\,$\pm$\,3.5&--0.38\,$\pm$\,0.28&12.25\,$\pm$\,0.69&7.29\,$\pm$\,0.33&0.54\,$\pm$\,0.09&[5]\\
326.270+00.783&3.0\,$\pm$\,0.4&--2.56\,$\pm$\,0.58&15.79\,$\pm$\,1.07&7.33\,$\pm$\,0.33&0.67\,$\pm$\,0.05&[5]\\
326.643+00.514&3.0\,$\pm$\,0.4&--1.35\,$\pm$\,0.58&13.84\,$\pm$\,1.07&7.32\,$\pm$\,0.33&0.59\,$\pm$\,0.05&[5]\\
327.300--00.548&3.2\,$\pm$\,0.4&--1.99\,$\pm$\,0.42&12.92\,$\pm$\,0.89&6.1\,$\pm$\,0.36&0.55\,$\pm$\,0.04&[5]\\
327.991--00.087&3.6\,$\pm$\,1.8&--0.79\,$\pm$\,0.35&10.84\,$\pm$\,0.81&6.0\,$\pm$\,0.36&0.47\,$\pm$\,0.05&[5]\\
328.572--00.527&3.4\,$\pm$\,0.4&--2.38\,$\pm$\,0.38&15.28\,$\pm$\,0.81&7.19\,$\pm$\,0.33&0.65\,$\pm$\,0.04&[5]\\
331.365+00.521&11.8\,$\pm$\,5.9&--3.3\,$\pm$\,0.52&12.93\,$\pm$\,1.0&4.8\,$\pm$\,0.34&0.85\,$\pm$\,0.33&[5]\\
332.145--00.452&3.7\,$\pm$\,0.4&--1.64\,$\pm$\,0.28&13.87\,$\pm$\,0.68&7.05\,$\pm$\,0.32&0.59\,$\pm$\,0.03&[5]\\
332.657--00.622&3.3\,$\pm$\,0.4&--2.1\,$\pm$\,0.78&14.77\,$\pm$\,1.34&7.15\,$\pm$\,0.32&0.62\,$\pm$\,0.06&[5]\\
332.762--00.595&3.8\,$\pm$\,0.4&--2.12\,$\pm$\,0.78&14.58\,$\pm$\,1.34&7.01\,$\pm$\,0.32&0.62\,$\pm$\,0.06&[5]\\
332.978+00.773&3.8\,$\pm$\,0.5&--2.25\,$\pm$\,0.33&9.98\,$\pm$\,0.76&4.0\,$\pm$\,0.35&0.43\,$\pm$\,0.03&[5]\\
333.011--00.441&3.6\,$\pm$\,0.4&--1.89\,$\pm$\,0.47&14.29\,$\pm$\,0.9&7.06\,$\pm$\,0.32&0.61\,$\pm$\,0.04&[5]\\
333.093+01.966&1.6\,$\pm$\,0.6&--1.13\,$\pm$\,0.48&14.05\,$\pm$\,0.95&7.67\,$\pm$\,0.35&0.55\,$\pm$\,0.04&[5]\\
333.627--00.199&3.2\,$\pm$\,0.4&--3.51\,$\pm$\,0.54&17.03\,$\pm$\,1.0&7.16\,$\pm$\,0.32&0.71\,$\pm$\,0.04&[5]\\
337.957--00.474&3.1\,$\pm$\,1.6&--1.13\,$\pm$\,0.38&10.74\,$\pm$\,0.83&5.6\,$\pm$\,0.35&0.44\,$\pm$\,0.04&[5]\\
338.706+00.645&4.3\,$\pm$\,0.4&--2.5\,$\pm$\,0.59&14.78\,$\pm$\,1.07&6.76\,$\pm$\,0.31&0.63\,$\pm$\,0.05&[5]\\
338.911+00.615&4.4\,$\pm$\,0.4&--2.67\,$\pm$\,0.59&15.01\,$\pm$\,1.07&6.73\,$\pm$\,0.31&0.64\,$\pm$\,0.05&[5]\\
338.934--00.067&3.2\,$\pm$\,0.4&--2.36\,$\pm$\,0.42&15.1\,$\pm$\,0.85&7.1\,$\pm$\,0.32&0.62\,$\pm$\,0.04&[5]\\
339.109--00.233&6.5\,$\pm$\,3.3&--2.05\,$\pm$\,0.58&9.98\,$\pm$\,1.06&4.2\,$\pm$\,0.32&0.47\,$\pm$\,0.09&[5]\\
339.134--00.377&3.0\,$\pm$\,0.4&--3.47\,$\pm$\,0.58&16.97\,$\pm$\,1.06&7.16\,$\pm$\,0.32&0.69\,$\pm$\,0.04&[5]\\
340.216+00.424&4.4\,$\pm$\,2.2&--2.77\,$\pm$\,0.59&12.09\,$\pm$\,1.08&4.8\,$\pm$\,0.33&0.52\,$\pm$\,0.07&[5]\\
340.678--01.049&2.3\,$\pm$\,2.3&--2.25\,$\pm$\,0.57&15.37\,$\pm$\,1.05&7.38\,$\pm$\,0.33&0.6\,$\pm$\,0.07&[5]\\
340.780--01.022&2.3\,$\pm$\,0.6&--2.6\,$\pm$\,0.57&15.93\,$\pm$\,1.05&7.38\,$\pm$\,0.33&0.62\,$\pm$\,0.04&[5]\\
340.862--00.870&2.3\,$\pm$\,2.3&--1.47\,$\pm$\,0.47&14.13\,$\pm$\,0.91&7.38\,$\pm$\,0.33&0.55\,$\pm$\,0.06&[5]\\
341.090--00.017&3.2\,$\pm$\,3.2&--2.75\,$\pm$\,0.3&15.68\,$\pm$\,0.7&7.07\,$\pm$\,0.32&0.64\,$\pm$\,0.09&[5]\\
342.277+00.311&9.6\,$\pm$\,4.8&--3.22\,$\pm$\,0.66&11.37\,$\pm$\,1.18&3.9\,$\pm$\,0.32&0.62\,$\pm$\,0.17&[5]\\
343.480--00.043&13.4\,$\pm$\,7.4&--2.22\,$\pm$\,0.42&16.48\,$\pm$\,0.88&8.1\,$\pm$\,0.35&1.13\,$\pm$\,0.57&[5]\\
343.914--00.646&2.8\,$\pm$\,1.4&--0.94\,$\pm$\,0.33&13.0\,$\pm$\,0.76&7.2\,$\pm$\,0.35&0.52\,$\pm$\,0.04&[5]\\
345.094--00.779&2.1\,$\pm$\,2.1&--3.59\,$\pm$\,0.62&17.59\,$\pm$\,1.13&7.43\,$\pm$\,0.33&0.68\,$\pm$\,0.07&[5]\\
345.202+01.027&1.1\,$\pm$\,0.6&--4.13\,$\pm$\,1.22&14.26\,$\pm$\,1.95&4.8\,$\pm$\,0.12&0.53\,$\pm$\,0.07&[5]\\
345.235+01.408&8.0\,$\pm$\,4.0&--1.94\,$\pm$\,0.68&12.68\,$\pm$\,1.22&6.0\,$\pm$\,0.35&0.63\,$\pm$\,0.14&[5]\\
345.410--00.953&2.6\,$\pm$\,0.6&--2.26\,$\pm$\,0.81&14.72\,$\pm$\,1.29&6.96\,$\pm$\,0.05&0.58\,$\pm$\,0.05&[5]\\
348.261+00.485&1.8\,$\pm$\,1.8&--3.74\,$\pm$\,0.64&17.99\,$\pm$\,1.15&7.53\,$\pm$\,0.34&0.68\,$\pm$\,0.06&[5]\\
348.691--00.826&3.4\,$\pm$\,0.3&--1.24\,$\pm$\,0.66&9.64\,$\pm$\,1.92&4.8\,$\pm$\,1.0&0.39\,$\pm$\,0.08&[5]\\
348.710--01.044&3.4\,$\pm$\,0.3&--1.97\,$\pm$\,0.57&13.04\,$\pm$\,1.83&6.2\,$\pm$\,1.0&0.52\,$\pm$\,0.07&[5]\\
350.991--00.532&13.7\,$\pm$\,6.9&--1.62\,$\pm$\,0.62&12.33\,$\pm$\,1.14&6.1\,$\pm$\,0.35&0.84\,$\pm$\,0.4&[5]\\
350.995+00.654&0.6\,$\pm$\,0.3&--6.76\,$\pm$\,1.34&27.67\,$\pm$\,2.21&10.57\,$\pm$\,0.34&1.0\,$\pm$\,0.08&[5]\\
351.130+00.449&1.4\,$\pm$\,0.7&--7.62\,$\pm$\,1.64&22.78\,$\pm$\,2.62&6.65\,$\pm$\,0.07&0.85\,$\pm$\,0.1&[5]\\
351.311+00.663&1.3\,$\pm$\,0.1&--6.84\,$\pm$\,0.69&23.23\,$\pm$\,1.24&7.71\,$\pm$\,0.35&0.86\,$\pm$\,0.05&[5]\\
351.383+00.737&1.3\,$\pm$\,0.1&--7.43\,$\pm$\,0.68&27.35\,$\pm$\,1.09&9.7\,$\pm$\,0.09&1.01\,$\pm$\,0.04&[5]\\
351.516--00.540&3.3\,$\pm$\,3.3&--3.9\,$\pm$\,0.68&15.33\,$\pm$\,1.93&5.7\,$\pm$\,1.0&0.61\,$\pm$\,0.11&[5]\\
351.688--01.169&14.2\,$\pm$\,1.0&--3.05\,$\pm$\,0.49&15.23\,$\pm$\,0.86&6.49\,$\pm$\,0.21&1.07\,$\pm$\,0.1&[5]\\
353.038+00.581&1.1\,$\pm$\,1.1&--5.39\,$\pm$\,1.12&21.03\,$\pm$\,1.87&7.78\,$\pm$\,0.35&0.77\,$\pm$\,0.08&[5]\\
353.076+00.287&0.7\,$\pm$\,1.5&--6.72\,$\pm$\,1.5&19.33\,$\pm$\,2.4&5.39\,$\pm$\,0.1&0.7\,$\pm$\,0.09&[5]\\
353.092+00.857&1.0\,$\pm$\,2.0&--5.72\,$\pm$\,1.12&20.47\,$\pm$\,1.89&7.1\,$\pm$\,0.4&0.75\,$\pm$\,0.09&[5]\\

\end{longtable}

\footnotetext[1]{The HII region was observed in different fields. The weighted mean for this HII region is calculated from the different measurements and is given in this table. This weighted mean was also used in our modeling effort.}

\footnotetext[2]{Observed values not given in original paper; values calculated here.}
\footnotetext[3]{Uncertainty absent in paper, as discussed in Sect.~\ref{sec:building}; we adopt 1.0\,$\times\,10^3$~K.}
\footnotetext[4]{Electron temperatures updated with values from \citet{Azcarate1990}.}

\footnotetext[5]{No error on the distance was provided by paper 2 or references therein. A 10\% uncertainty was assumed.}
\footnotetext[6]{Distance as calculated from the radial velocities in \citet{Paladini2003} different from cited distance in paper 3. The new calculated value was adopted.}
}}

\end{document}